%% file: bare_jrnl.tex
\documentclass[journal]{IEEEtran}

\ifCLASSINFOpdf
\else
   \usepackage[dvips]{graphicx}
\fi

\usepackage{algorithmic}
\usepackage{graphicx}
\usepackage{textcomp}
\usepackage{microtype}
\usepackage{cite}
\usepackage{amsmath,amssymb,amsfonts}
\usepackage{algorithmic}
\usepackage{textcomp}
\usepackage{multirow}
\usepackage{comment}
\usepackage{booktabs, soul}
\usepackage{tabularx, color, soul}
\usepackage{pifont}
\usepackage[table]{xcolor}
\usepackage{diagbox}
\usepackage{slashbox}
\usepackage{array, makecell}
\usepackage{commath}

\setstcolor{red}
\makeatletter
\newcommand\notsotiny{\@setfontsize\notsotiny{6}{7}}
\makeatother
\newcolumntype{x}[1]{>{\centering\arraybackslash\hspace{0pt}}p{#1}}

\usepackage[acronym,shortcuts,acronymlists={hidden}]{glossaries}
\usepackage[colorlinks]{hyperref}       
\input{glossary.tex}
\def\BibTeX{{\rm B\kern-.05em{\sc i\kern-.025em b}\kern-.08em
    T\kern-.1667em\lower.7ex\hbox{E}\kern-.125emX}}
    
\newcolumntype{C}{>{\centering\arraybackslash}X}
\usepackage{caption,subcaption,graphicx}
\usepackage{pifont}

\definecolor{redtable}{RGB}{255, 153, 153}
\definecolor{greentable}{RGB}{128, 255, 128}

\begin{document}

\title{Perceptual Quality Assessment of HEVC and VVC Standards for 8K Video}

\author{Charles Bonnineau, Wassim Hamidouche, \IEEEmembership{Member, IEEE}, \\ Jérôme Fournier,  Naty Sidaty, \IEEEmembership{Member, IEEE}, Jean-François Travers and Olivier Déforges\vspace{-0.1in}
\thanks{C. Bonnineau, W. Hamidouche and J. Fournier are with the Institute of Research and Technology (IRT) b$<>$com, 35510 Cesson Sévigné, France, e-mail: (\href{mailto:charles.bonnineau@b-com.com}{charles.bonnineau@b-com.com})}
\thanks{C. Bonnineau, W. Hamidouche and O. Déforges are also with Univ. Rennes, INSA Rennes, CNRS, IETR - UMR 6164, 20 Avenue des Buttes de Coesmes, 35708 Rennes, France, e-mail: (\href{mailto:whamidou@insa-rennes.fr}{whamidou@insa-rennes.fr}).}
\thanks{C. Bonnineau, N. Sidaty, J-F. Travers are also with TDF, 35510 Cesson-Sévigné, France.}
\thanks{J. Fournier is also with Orange Labs, 35510 Cesson-Sévigné, France. }}

\markboth{IEEE Transactions on Broadcasting - accepted version}
{Shell \MakeLowercase{\textit{et al.}}: Bare Demo of IEEEtran.cls for IEEE Journals}

\maketitle


\begin{abstract}


With the growing data consumption of emerging video applications and users' requirement for higher resolutions, up to 8K, a huge effort has been made in video compression technologies. Recently, \gls{VVC} has been standardized by the \gls{MPEG}, providing a significant improvement in compression performance over its predecessor \gls{HEVC}. In this paper, we provide a comparative subjective quality evaluation between \gls{VVC} and \gls{HEVC} standards for 8K resolution videos. In addition, we evaluate the perceived quality improvement offered by 8K over UHD 4K resolution. The compression performance of both \gls{VVC} and \gls{HEVC} standards has been conducted in \gls{RA} coding configuration, using their respective reference software, \gls{VTM} and \gls{HM}. Objective measurements, using {PSNR}, {MS-SSIM} and {VMAF} metrics have shown that the bitrate gains offered by \gls{VVC} over \gls{HEVC} for 8K video content are around  31\%,  26\%  and  35\%, respectively. Subjectively, \gls{VVC} offers an average of around 41\% of bitrate reduction over \gls{HEVC} for the same visual quality. A compression gain of 50\% has been reached for some tested video sequences regarding a Student's t-test analysis. In addition, for most tested scenes, a significant visual difference between uncompressed 4K and 8K has been noticed.

\end{abstract}

\vspace{-0.1in}

\IEEEpeerreviewmaketitle

\glsresetall

\section{Introduction}


\IEEEPARstart{W}{ith} the latest \gls{UHDTV} system~\cite{itu-r_recommendation_BT2020-1} deployment, the \gls{QoE} of users is expected to improve by introducing new features to the existing \gls{HDTV} system~\cite{itu-r_recommendation_BT709-5}, including \gls{HDR}, wider color gamut, \gls{HFR}, and higher spatial resolutions, with 4K (3840 $\times$ 2160) and 8K (7680 $\times$ 4320) \cite{nilsson2015ultra, sugawara2012uhdtv}. The delivery of these video formats on current broadcast infrastructures is a real challenge and requires efficient compression methods to reach the available bandwidth while ensuring a higher video quality.


\begin{figure*}[t]
\begin{minipage}[b]{0.33\linewidth}
  \centering
  \centerline{\includegraphics[width=1\linewidth]{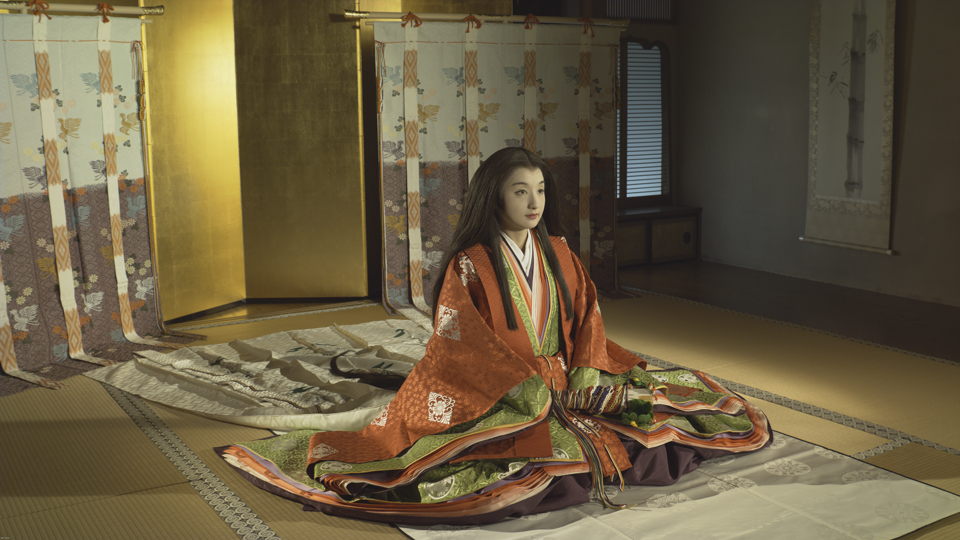}}
  \centerline{(a) LayeredKimono}\medskip
\end{minipage}
\begin{minipage}[b]{0.33\linewidth}
  \centering
  \centerline{\includegraphics[width=1\linewidth]{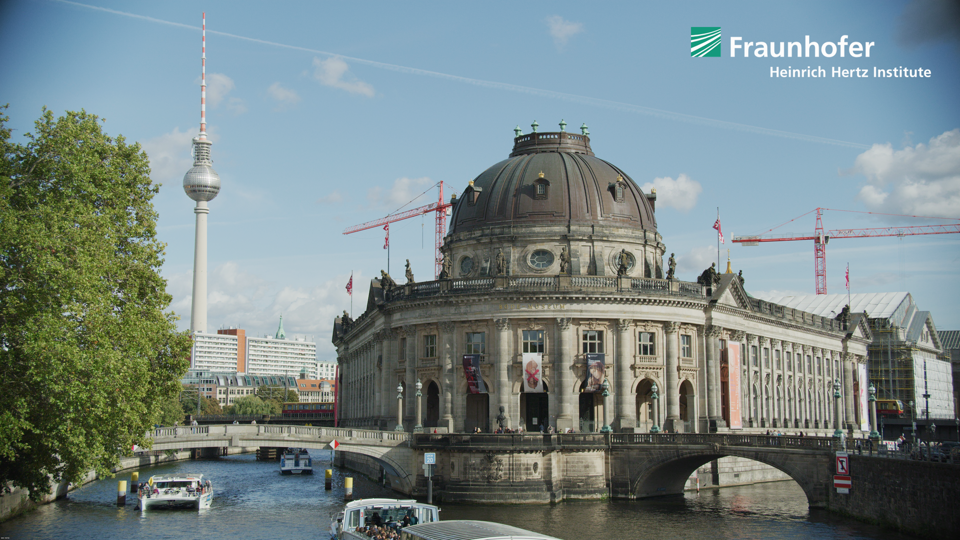}}
  \centerline{(b) BodeMuseum}\medskip
\end{minipage}
\hfill
\begin{minipage}[b]{0.33\linewidth}
  \centering
  \centerline{\includegraphics[width=1\linewidth]{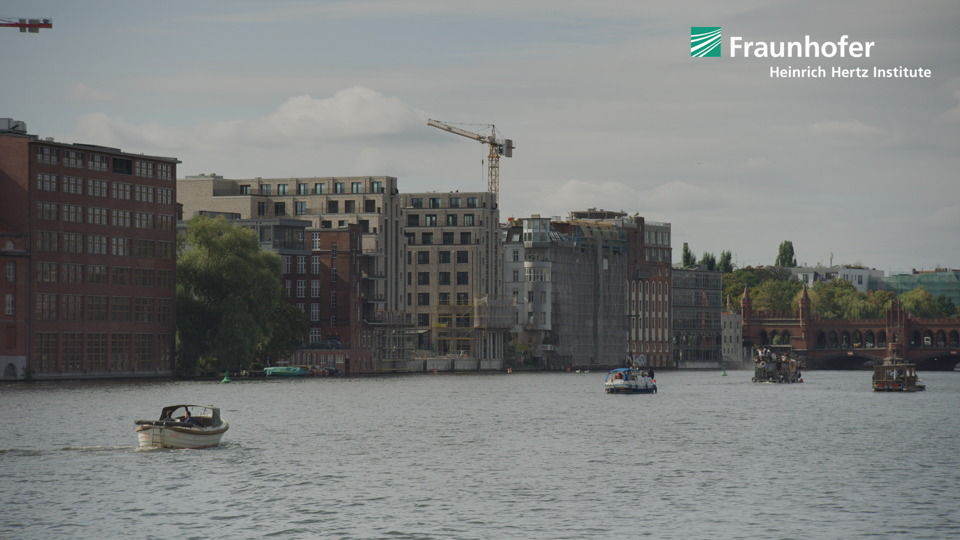}}
  \centerline{(c) OberbaumSpree}\medskip
\end{minipage}
\hfill
\begin{minipage}[b]{0.33\linewidth}
  \centering
  \centerline{\includegraphics[width=1\linewidth]{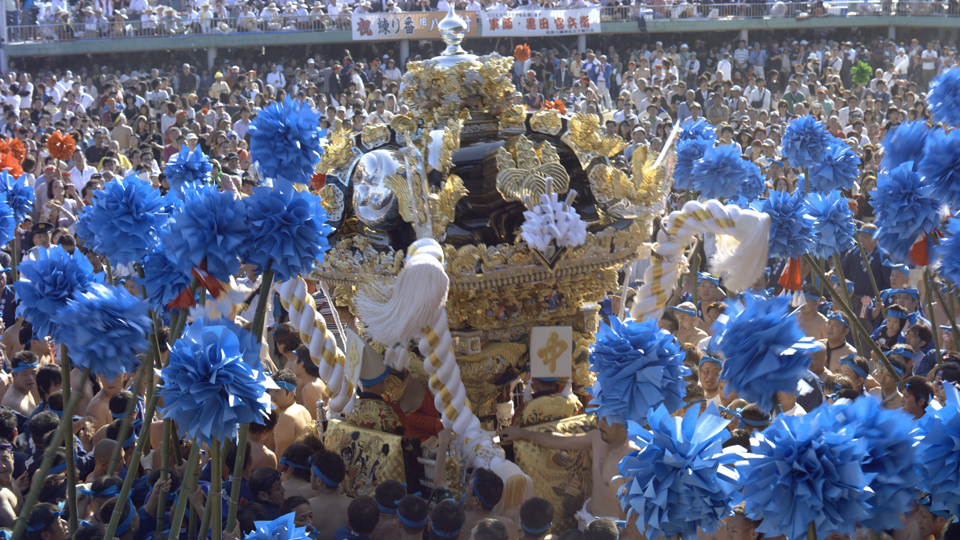}}
  \centerline{(d) Festival2}\medskip
\end{minipage}
\begin{minipage}[b]{0.33\linewidth}
  \centering
  \centerline{\includegraphics[width=1\linewidth]{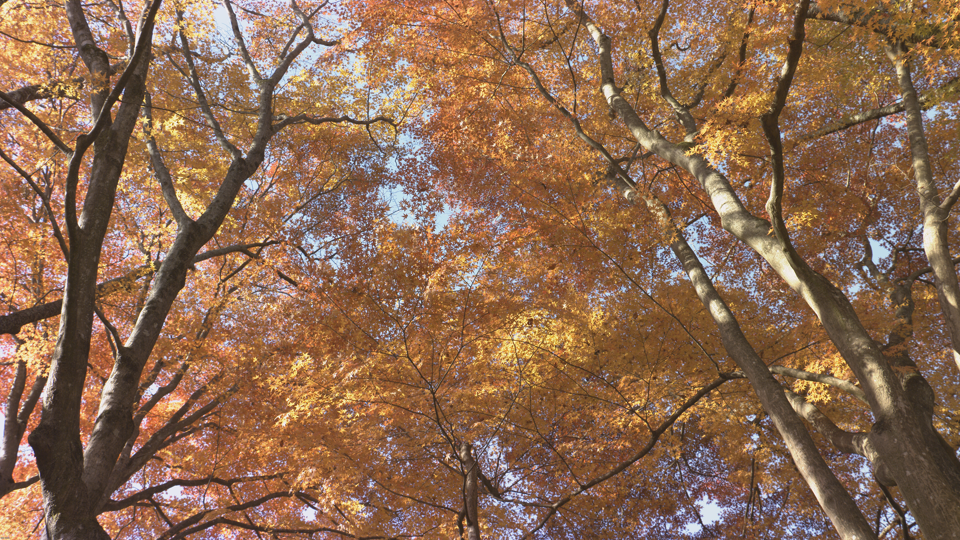}}
  \centerline{(e) JapaneseMaple}\medskip
\end{minipage}
\hfill
\begin{minipage}[b]{0.33\linewidth}
  \centering
  \centerline{\includegraphics[width=1\linewidth]{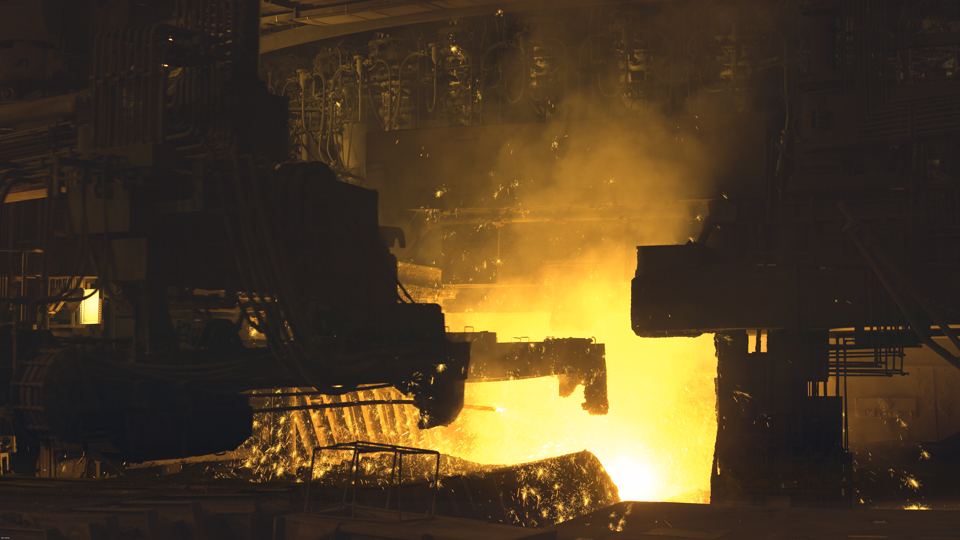}}
  \centerline{(f) SteelPlant}\medskip
\end{minipage}
\hfill
\vspace{-0.2in}
\caption{Snapshots of the six selected 8K test video sequences.}
\vspace{-0.1in}
\label{fig:sequence_screenshot}
\end{figure*}

Contributions to video coding standards like \gls{HEVC}~\cite{sullivan2012overview} or its successor \gls{VVC}, finalized in July 2020 as ITU-T H.266 | MPEG-I - Part 3 (ISO/IEC 23090-3) standard~\cite{9328514, hamidouche2021versatile}, enable video signal compression to be continuously improved through the standardization bodies. Although \gls{HEVC} has brought a significant bitrate reduction for 4K delivery, its efficiency is not enough for 8K applications. Several studies have shown that the bitrate required by \textcolor{black}{\gls{HEVC}} for 8K applications in 60Hz and 120Hz (temporally scalable) is around 80Mbps \cite{sugito2020video, ichigaya2016required, iwasaki2020required}. In practice, an 8K 120Hz \gls{HEVC} codec \cite{sugito2015hevc, sugito2018study} has been used for Japan's satellite broadcasting by using DVBS2X~\cite{etsi_dvb-s2x}. In that case, the use of a complete transponder or multiple bonded transponders can reach bandwidth in the range 70-80Mbps. For terrestrial transmission, such bandwidth requirements prevent the deployment of more than one 8K \gls{HEVC} program per \gls{UHF} channel, as practical DVB-T2~\cite{etsi_dvb-t2} channels offer bandwidth in the range of 30-40Mbps over an 8MHz channel. \textcolor{black}{Thus, significant compression gains need to be achieved to ensure the successful deployment of 8K video services.}

This paper provides both subjective and objective quality assessments of the two latest \gls{MPEG} video coding standards for 8K video coding. We selected 8K sequences with various spatial and temporal characteristics to provide a fair evaluation. The compression points have been generated using the \glsfirst{RA} mode of the \gls{VVC} and \gls{HEVC} reference software models, called \gls{VTM} and \gls{HM}, respectively. For subjective quality assessment, we used the \glsfirst{DSCQS} method described in  Recommendation BT.500-14 \cite{itu-r_recommendation_BT500-14} standardized at ITU-R. This study includes \gls{RD} curves, \gls{BD} bitrate evaluation, and a Student's t-test, offering a robust statistical analysis. 

The contributions of this work are the following:

\begin{itemize}
    \item Assess the compression gain offered by \gls{VVC} over \gls{HEVC} for 8K video contents. This gain represents approximately 41\% of bitrate saving for the same visual quality,
    \item Determine the required bitrate for transparency, i.e., no visual difference is perceived between the source and decoded video,
    \item Confirm that non-expert viewers can see the difference between 4K and 8K resolutions and measure that difference,
    \item Evaluate several objective quality metrics based on the subjective test statistics collected on the 8K video dataset.
\end{itemize}

The rest of this paper is organized as follows. \textcolor{black}{Section \ref{sec:related_works} provides an overview of existing studies for 4K and 8K video quality assessment.} Section~\ref{sec:subj_quality} describes the subjective test materials, including the test sequences, the codecs configuration, and the subjective test methodology. The results of both the objective and the subjective experiments are given in Section~\ref{sec:results}. Finally, Section~\ref{sec:conclusion} concludes the paper.
\textcolor{black}{\section{Related works}
\label{sec:related_works}}



\textcolor{black}{Recently, a study was conducted to evaluate different scenarios for 8K video delivery with 4K backward compatibility relaying on objective quality metrics~\cite{bonnineau2020versatile}. It was shown that \gls{VVC} offers around 40\% of bitrate reduction over \gls{HEVC} for the same \gls{PSNR} quality on 8K video resolution~\cite{bonnineau2020ANOE}. Although recently developed objective quality metrics, like \gls{VMAF}~\cite{vmaf}, are more correlated to subjective test scores, it is acknowledged that these quality metrics still lack fidelity regarding the viewing conditions and the human visual system. Thus, rigorous perceptual quality assessment methodologies have been developed to fairly evaluate compression algorithms and ensure experiment reproducibility~\cite{itu-r_recommendation_BT500-13, sotelo2017subjective}.}





\textcolor{black}{For instance, Tan \textit{et al.}~\cite{tan2015video} have demonstrated that a difference of 15\% of compression gain is noticed depending on whether the objective or subjective quality is considered when evaluating \gls{HEVC} over \gls{AVC}. This evaluation has been conducted using the respective reference implementations of both standards for resolutions ranging from 480p to 2160p. Another perceptual study has confirmed that a bitrate saving in the range 55-87\% for the same perceived quality is enabled by \gls{HEVC} over \gls{AVC} on a bench of sequences, including 4K contents \cite{tabatabai2014compression}. Regarding VVC and HEVC comparison, a recent subjective test has validated that \gls{VVC} offers around 40\% or bitrate reduction for the same perceived quality targeting 4K and HD contents \cite{sidaty2019compression}. In addition to \gls{HEVC} and \gls{VVC}, subjective quality assessment of \gls{AV1} has been included in the work of Zhang \textit{et al.}~\cite{zhang2020comparing} for 4K video resolution. The results have shown that, at the same video bitrate level, \gls{AV1} and \gls{HM} are not significantly different in terms of perceived quality.}


\textcolor{black}{For 4K video resolution broadcasting with \gls{HEVC}, a study has been conducted regarding target bitrates in the range 18-36Mbps \cite{bae2013assessments}. This experiment has demonstrated that 4K resolution can reach a good perceptual quality at a bitrate of 18Mbps using \gls{HEVC}.}

\begin{table}[t]
		\caption{Parameters of the 8K test video sequences. All sequences are in 4:2:0 color sub-sampling format.}
		\label{tab:seq_detail}
		\begin{tabularx}{\linewidth}{@{}lx{1.4cm}CCCCC@{}}
			\midrule[0.3mm]
			\multirow{2}{*}{Sequence} & Resolution (W $\times$ H) & Frame-rate & Frames & Color space & Bitdepth & Src  \\
			\midrule[0.2mm]
			\it BodeMuseum& 7680$\times$4320 & 60fps & 600 & BT.709 & 10 & HHI\\
			\it OberbaumSpree& 7680$\times$4320 & 60fps & 600 & BT.709 & 10 & HHI\\
			\it LayeredKimono& 7680$\times$4320 & 60fps & 300 & BT.2020 & 10 & ITE\\
			\it Festival2& 7680$\times$4320 & 60fps & 300 & BT.2020 & 10 & ITE\\
			\it JapaneseMaple& 7680$\times$4320 & 60fps & 300 & BT.2020 & 10 & ITE\\
			\it SteelPlant& 7680$\times$4320 & 60fps & 600 & BT.2020 & 10 & ITE\\
           \midrule[0.3mm]
		\end{tabularx}
\vspace{-0.2in}
\end{table}

\textcolor{black}{Concerning 8K resolution videos, several studies have shown that the bitrate required for 8K applications is approximately 80Mbps using \gls{HEVC} \cite{sugito2020video, ichigaya2016required, iwasaki2020required}. The \gls{QoE} of 8K contents has also been assessed regarding different use-cases by using specific contents~\cite{shishikui2021quality}, e.g., food, people.}


\textcolor{black}{In this paper, we provide a subjective evaluation between \gls{HEVC} and \gls{VVC} for 8K resolution video. To the best of our knowledge, this is the first quality assessment study based on those two \gls{MPEG} standards for 8K. Also, we provide an analysis on the gain in terms of quality enhancement offered by 8K over 4K for the uncompressed selected contents.} 

\vspace{0.2in}
\section{Subjective quality assessment of 8K resolution}
\label{sec:subj_quality}

This section provides details regarding the test sequences, the subjective test settings, and the experimental environment.

\subsection{Test video sequences}
\label{sec:sequences}

\begin{figure}[t]
\centering
\includegraphics[width=\linewidth]{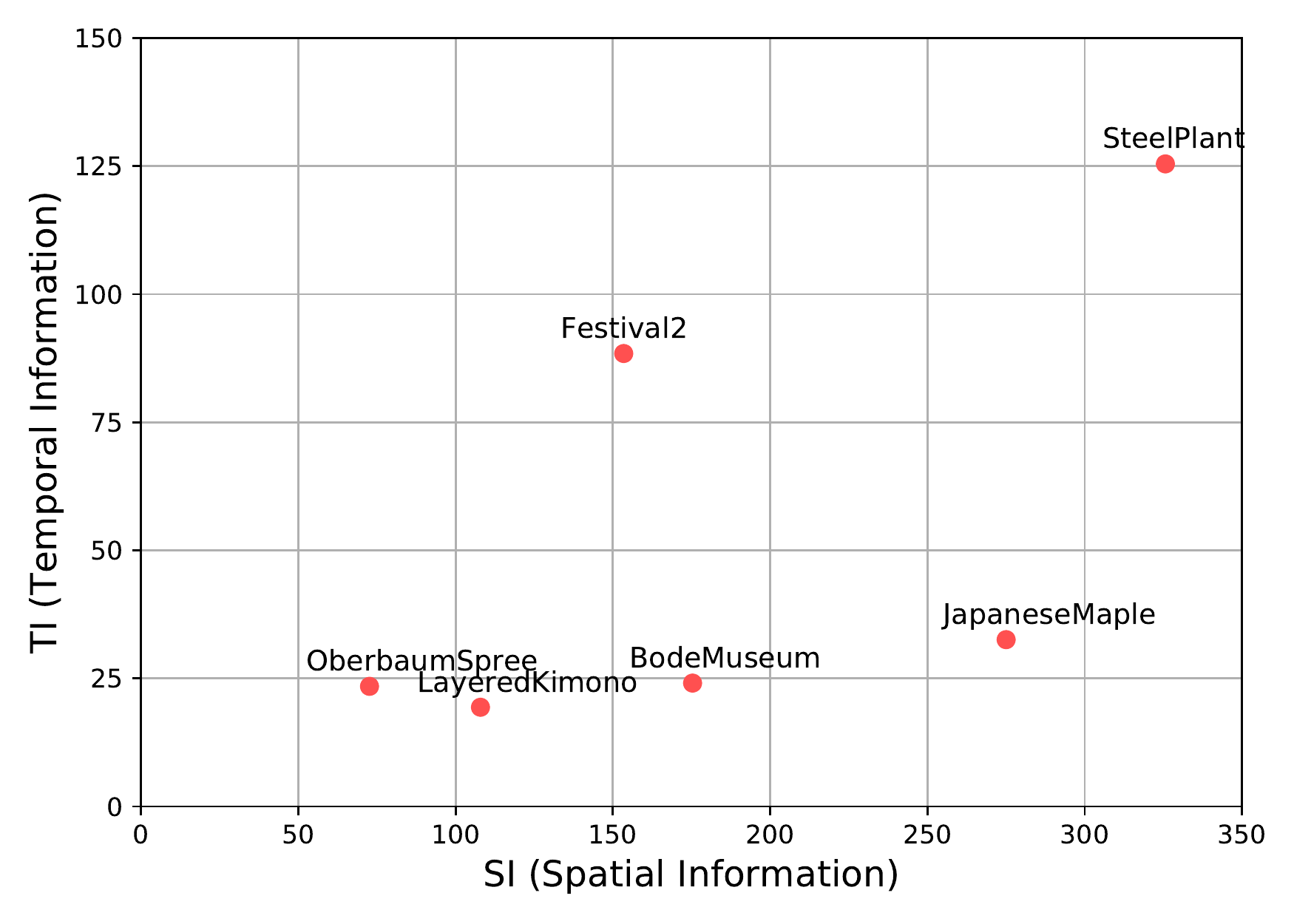}
\caption{SI-TI graph of the tested 8K video sequences.}
\label{fig:siti}
\vspace{-0.1in}
\end{figure}

In this study, we selected six test video sequences over multiple videos collected from the \gls{ITE}\footnote{\url{https://www.ite.or.jp/content/test-materials/}} and the \gls{HHI} \cite{jvet-q0791} 8K video databases. The scenes were chosen based on video features like color, movement, texture, and homogeneous content, leading to different behaviors of the compression algorithms. We also considered the relevance of the 8K resolution in the scene selection. The details of the 8K test sequences are reported in Table~\ref{tab:seq_detail}. Screenshots of the selected scenes are given in Fig. \ref{fig:sequence_screenshot}. To ensure homogeneity over video sequences and keep the same display parameters for the whole experiment, we performed a color space conversion from BT.709 \cite{itu-r_recommendation_BT709} to BT.2020 \cite{itu-r_recommendation_BT2020} for \textit{BodeMuseum} and \textit{OberbaumSpree} scenes. Also, as the sequences \textit{LayeredKimono}, \textit{Festival2}, and \textit{JapaneseMaple} contain fewer frames than the others, we played them back in mirror mode after 5 seconds to get 10 seconds videos while preserving the motion continuity of the scene. For those sequences, the motion direction change was coherent with the initial content. 

The spatial and temporal information (SI-TI) \cite{itu-r_recommendation_BT500-14} of the selected sequences is plotted in Fig. \ref{fig:siti}. This 2D plan shows that the contents selected for the study are diverse regarding spatio-temporal features.  

Based on these six uncompressed (raw) selected 8K video sequences (scenes), ten \glspl{PVS} are generated per scene:

\begin{itemize}
    \item one 8K (7680$\times$4320) hidden reference uncompressed video.
    \item one 4K (4320$\times$2160) uncompressed video. In that case, the source signal is first downscaled to 4K and then rescaled to 8K by using the \textit{Lanczos3}~\cite{duchon1979lanczos} filter provided by \textit{ffmpeg}\footnote{\url{https://www.ffmpeg.org/}} for both operations.
    \item 8K video encoded at four bitrates with \gls{HEVC}.
    \item 8K video encoded at four bitrates with \gls{VVC}.
\end{itemize}

In total, 60 video sequences are evaluated in this study.


\begin{table}[t]
		\caption{Selected \acrshort{QP} and corresponding bitrates (Mbps), for both \gls{VTM} and \gls{HM} codecs, according to the test sequence.}
		\label{tab:qp_rate}
		\begin{tabularx}{\linewidth}{@{}lCCCCC@{}}
			\midrule[0.3mm]
			\multirow{2}{*}{Sequence}& \multirow{2}{*}{Codec} & $R_1$ (QP/Mbps) & $R_2$ (QP/Mbps) & $R_3$ (QP/Mbps) & $R_4$ (QP/Mbps) \\
			\midrule[0.2mm]
			\multirow{2}{*}{\it LayeredKimono}& \gls{HEVC} & 38/1.9 & 34/3.2 & 29/6.3 & 26/11.4 \\
			& \gls{VVC} & 37/1.8 & 32/3.4 & 27/6.5 & 24/10.8 \\
			\midrule[0.2mm]
			\multirow{2}{*}{\it BodeMuseum}& \gls{HEVC} & 38/4.7 & 33/9.8 & 28/22.5 & 25/45.4 \\
			& \gls{VVC} & 37/4.8 & 32/10.1 & 27/22.6 & 24/42.9\\
			\midrule[0.2mm]
			\multirow{2}{*}{\it OberbaumSpree}& \gls{HEVC} & 38/3.3 & 33/7.4 & 28/17.5 & 24/40.5\\
			& \gls{VVC} & 37/3.6 & 32/8.1 & 27/18.6 & 23/43.9\\
			\midrule[0.2mm]
			\multirow{2}{*}{\it Festival2}& \gls{HEVC} & 39/17.5 & 34/32.1 & 29/59.5 & 24/130.4\\
			& \gls{VVC} & 37/17.4 & 32/32.2 & 27/61.1 & 22/135.5\\
			\midrule[0.2mm]
			\multirow{2}{*}{\it JapaneseMaple}& \gls{HEVC} & 43/15.2 & 38/34.9 & 33/76.1 & 28/168\\
			& \gls{VVC} & 42/15.9 & 37/35.7 & 32/79.8 & 27/174.9\\
			\midrule[0.2mm]
			\multirow{2}{*}{\it SteelPlant}& \gls{HEVC} & 42/19.6 & 38/40.5 & 33/86.9 & 28/175.5\\
			& \gls{VVC} & 42/18.0 & 37/42.9 & 32/91.1 & 27/180.5\\
			\midrule[0.3mm]
		\end{tabularx}
\end{table}

The Common Test Conditions for \gls{VTM} \cite{reference_software_vvc} and \gls{HM} \cite{reference_software_hevc} in \gls{RA} coding mode \textcolor{black}{for main10 profile} were used to perform a fair rate/distortion evaluation. These software models provide a reference implementation of the compression standards, representing their upper-bound coding performance with a moderate optimization level. \textcolor{black}{For both codec, a GOP size of 16 and an Intra Period of 64 frames were used.} For each scene, the test points are obtained using different fixed \glsfirst{QP} values. To cover a wide range of visual quality, we determined the highest bitrate value considering the transparency, i.e., the bitrate for which degradation starts to appear, as the highest bitrate point for each sequence. Also, the bitrates were carefully selected so that each bitrate $R_i$ is approximately half of the next bitrate $R_{i+1}$ and each \gls{VVC} bitrate $R_i^{VVC}$ is equal to the corresponding HEVC bitrate $R_i^{HEVC}$ for $i \in \{1, 2, 3, 4\}$. The used QPs and bitrates for each sequence are given in Table \ref{tab:qp_rate}. We can note that the bitrate selected for transparency varies from 11Mbps to 180Mbps, depending on the test sequence.

\subsection{Subjective testing procedure}

\begin{figure}[t]
\centering
\includegraphics[width=\linewidth]{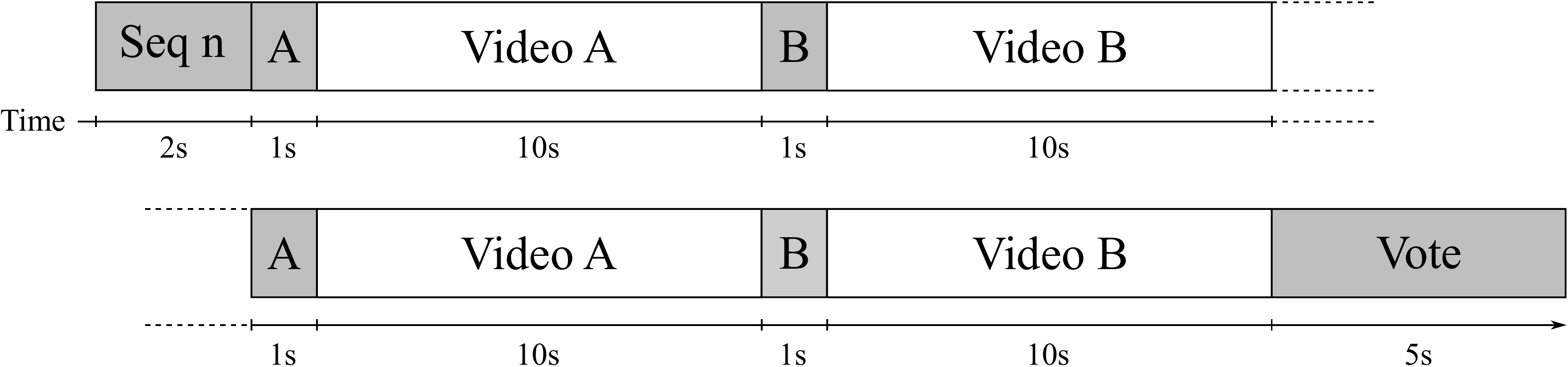}
\caption{Subjective \glsfirst{BTC} structure according to the \acrshort{DSCQS} evaluation methodology.}
\label{fig:dscqs}
\vspace{-0.1in}
\end{figure}

In this study, we used the method described in the ITU-R Recommendation BT.500-14 \cite{itu-r_recommendation_BT500-14}, called \gls{DSCQS}, to collect the video quality scores from participants. This testing method requires a prior pseudo-random sequencing of the testing videos, as the observer has no interactivity with the player. Thus, each test session of the \gls{DSCQS} method consists of different random series of \glspl{BTC} presentations. This method presents the test videos by pairs ("video A" and "video B") separated with annotated mid-greys. For each \gls{BTC}, both "video A" and "video B" are repeated twice. An example of \gls{BTC} used for evaluation is illustrated in Fig. \ref{fig:dscqs}. Each presented pair contains the implicit 8K uncompressed reference and one random \gls{PVS} over all the ten configurations, i.e., the same scene encoded with \gls{HEVC} or \gls{VVC} at four bitrates or the uncompressed sequence in 4K or 8K resolution. Also, to prevent visual fatigue, the test is divided into three sessions of 20 minutes each. Before each experiment, participants receive clear explanations about the evaluation procedures.

After the first "video A/video B" pair presentation, the participant could report his opinion about the perceived video quality on two vertical lines with the corresponding sequence index for both "video A" and "video B". For this testing method, the vertical rating lines are divided into five segments of the same height and scaled from the lower to the higher quality with the labels \textit{Bad}, \textit{Poor}, \textit{Fair}, \textit{Good}, and \textit{Excellent}. After each video pair visualization, participants can vote by annotating both videos along the continuous quality scale. The scores are then collected by converting the annotations into a value between 0 and 100.

\subsection{Experimental environment}

\begin{table}[t]
		\caption{Test logistics.}
		\label{tab:setup_settings}
		\begin{tabularx}{\linewidth}{@{}C|C@{}}
			\midrule[0.3mm]
			Monitor &  SONY 85'' KD-85ZG \\
			Player & Zaxel's Zaxtar 5 8K \\
			Peak luminance & 120 cd/m$^2$ \\
			\midrule[0.2mm]
			Video Format & 7680x4320/60p/YUV4:2:0/10bits\\
			Viewing distance & 0.8H (approximtely 0.8m)\\
			\midrule[0.2mm]
			Background color & D65 mid-grey \\
			\multirow{2}{*}{Background luminance} & 15\% of the screen maximum luminance\\
			\midrule[0.3mm]
		\end{tabularx}
\vspace{-0.1in}
\end{table}

\begin{figure}[t]
\centering
\includegraphics[width=\linewidth]{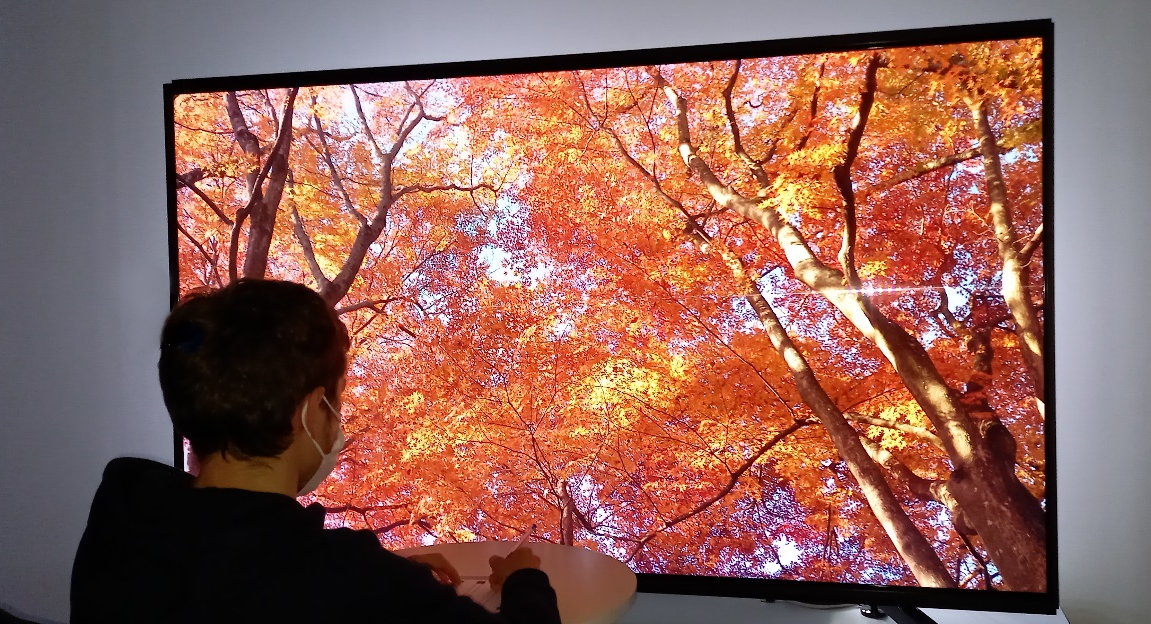}
\caption{Illustration of the laboratory environment, compliant with the ITU-R BT500-13 Recommendation~\cite{itu-r_recommendation_BT500-13}.}
\label{fig:subj_cond}
\vspace{-0.2in}
\end{figure}

This subjective study has been conducted in a controlled laboratory environment that follows the ITU-R Rec. BT.500-14~\cite{itu-r_recommendation_BT500-14}. The objective is to offer visualization comfort to participants and ensure the reproducibility of the test. All the experimental setup details are reported in Table \ref{tab:setup_settings}. A picture illustrating the test conditions is given in Fig.~\ref{fig:subj_cond}. A total of 22 non-expert observers aged from 22 to 53 years have taken part in this experiment. All participants have been screened for normal visual acuity and color blindness using the Ishihara and Snellen vision tests, as described in the ITU-R Recommendation BT.500-14 \cite{itu-r_recommendation_BT500-14}. \textcolor{black}{To detect outliers, the rejection method based on the Kurtosis coefficient} from this same recommendation has been applied and has validated the overall participant's reported votes.

\subsection{Subjective quality assessment}
\label{sec:subj_quality_assesssment}

At the end of the subjective test sessions, the results for each scene are assessed by the \gls{DMOS}, corresponding to the average of the difference between the hidden reference and the corresponding \gls{PVS} scores computed by: 

\begin{equation}
    \bar{x}_a = \frac{1}{n} \, \sum_{i=1}^{n} x_{i,a}, 
\end{equation}

where $n$ is the total number of valid participants, $\bar{x}_a$ is the \gls{DMOS} value of the tested configuration $a$, $a \in \{R^m_j$, 4K, 8K (ref)\} for $j \in \{1,2, 3, 4\}$ and $m \in \{VVC, HEVC\}$ and $x_{i,a}$ is the differential score computed as:

\begin{equation}
    x_{i,a} = 100 - (y_{i,ref} - y_{i,a}),
\end{equation}

with the pair $(y_{i,ref}, y_{i, a})$ representing the scores attributed by the participant $i$, $i \in \{1, \dots ,n\}$, to respectively the hidden reference (8K) and the tested configuration $a$, i.e.  both  videos  of a  given \gls{BTC}.

To ensure that the vote distributions are normal, the bias reduction technique described in the ITU-T P.913 Recommendation \cite{itu-r_recommendation_BT913} has been applied. Thus, from each resulting \gls{DMOS} $\bar{x}_a$, the associated confidence intervals at 95\% $(\bar{x}_a - c_a, \bar{x}_a + c_a)$ can be computed as follows:

\begin{equation}
    c_a = 1.96 \, \frac{s_a}{\sqrt{n}},
\end{equation}

where $s_a$ is the standard deviation of the tested configuration $a$ computed as:

\begin{equation}
    s_a=\sqrt{\sum_{i=1}^{n}\frac{ \left(x_{i,a}-\bar{x}_a \right )^2}{(n-1)}},
\end{equation}

with $x_{i,a}$ and $\bar{x}_{a}$ corresponding to the differential score of the observer $i$, $i \in \{1, \dots ,n\}$, and the \gls{DMOS} score of the tested configuration $a$, respectively.

In addition, a Student's t-test with a two-tailed distribution is performed to provide a more rigorous analysis. More details are given in Section~\ref{sec:suj_quality}

\section{Experimental results}
\label{sec:results}

This section presents and discusses the results of both objective and subjective evaluation scores. An assessment of the objective metrics performance compared to the subjective scores for 8K video contents is also investigated. 

\subsection{Objective results}
\label{sec:obj_study}

\begin{figure*}[t]
\small
\begin{minipage}[b]{0.32\linewidth}
  \centering
  \centerline{\includegraphics[width=1\linewidth]{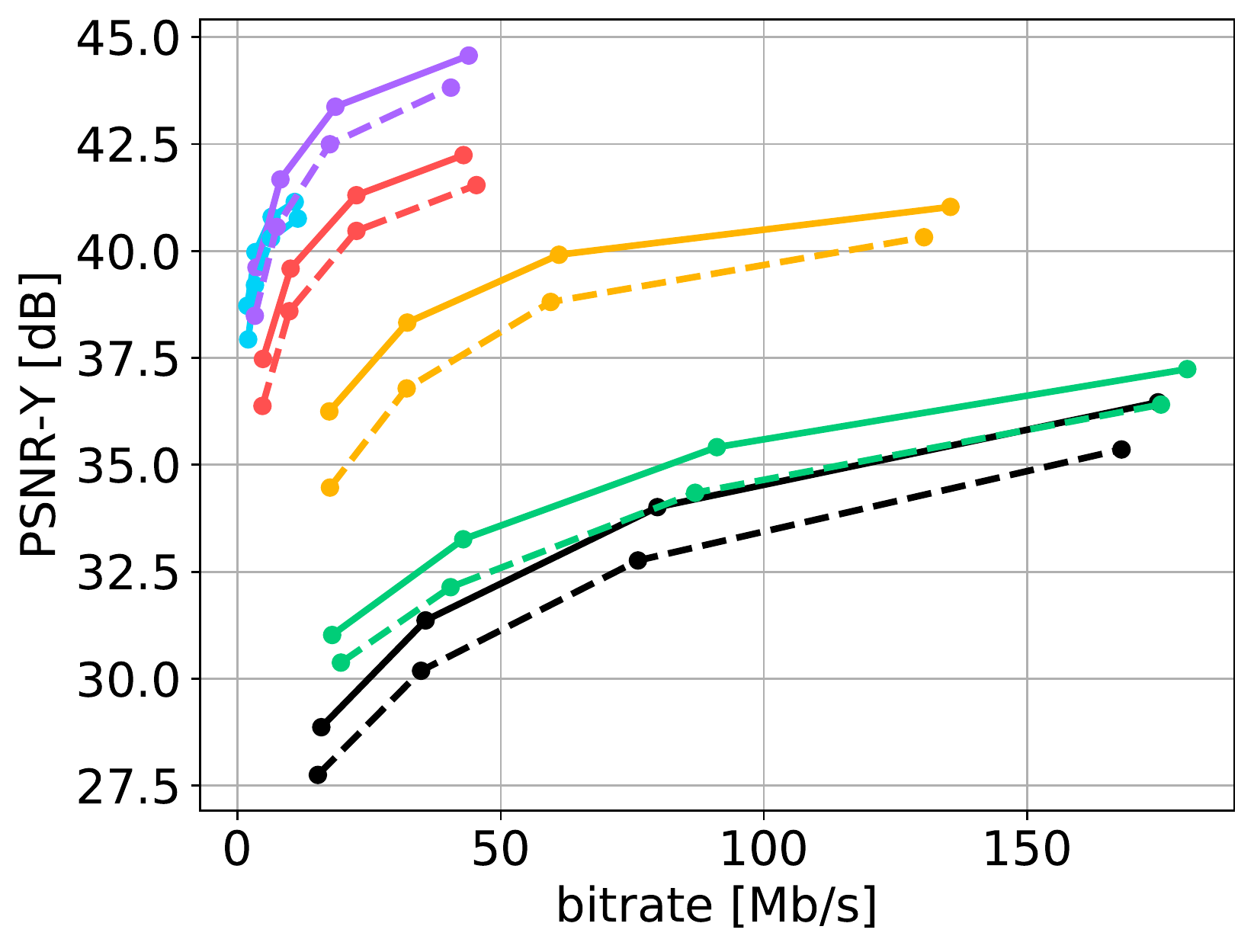}}
  \centerline{(a) PSNR}\medskip
\vspace{-0.1in}
\end{minipage}
\begin{minipage}[b]{0.32\linewidth}
  \centering
  \centerline{\includegraphics[width=1\linewidth]{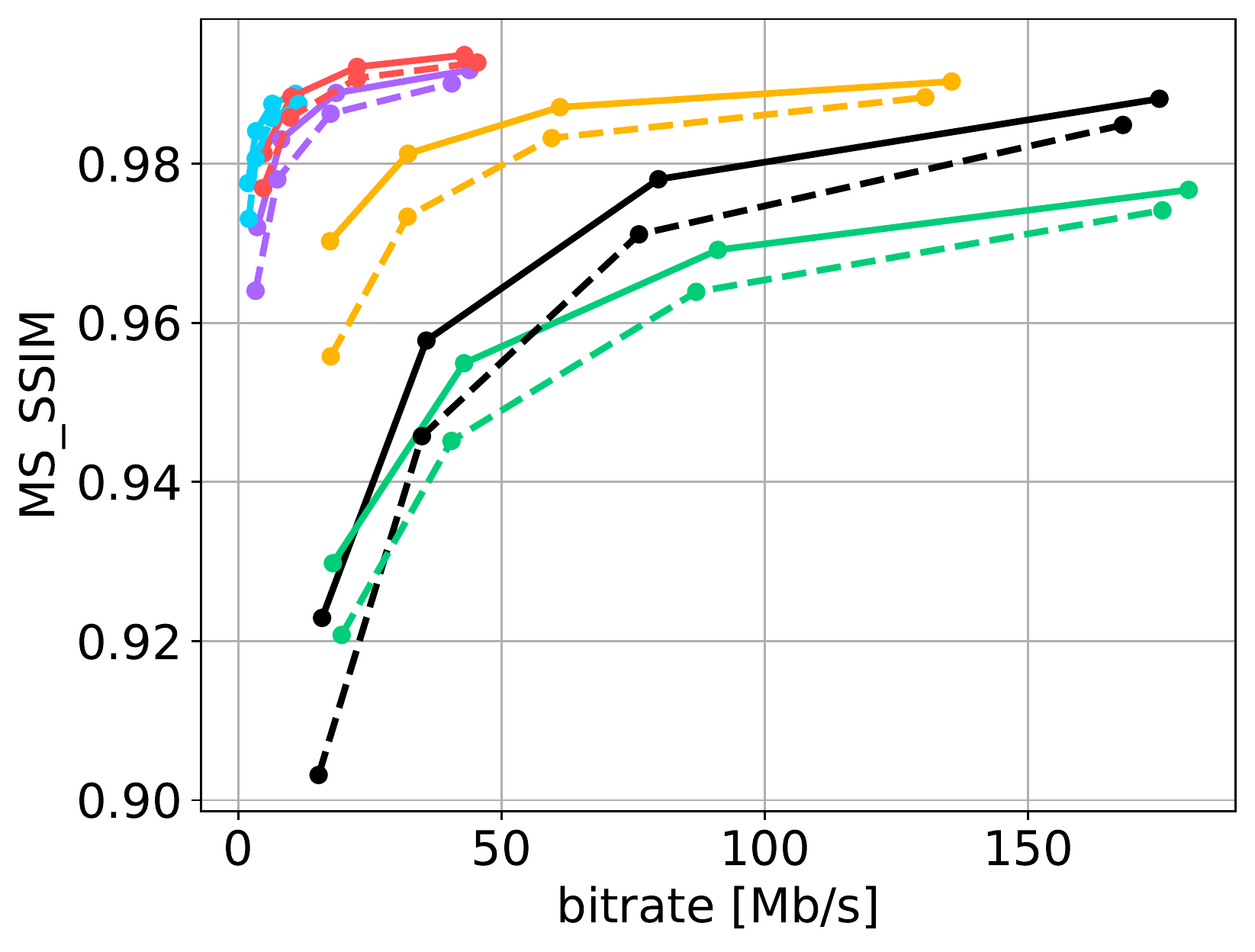}}
  \centerline{(b) {MS-SSIM}}\medskip
\vspace{-0.1in}
\end{minipage}
\hfill
\begin{minipage}[b]{0.32\linewidth}
  \centering
  \centerline{\includegraphics[width=1\linewidth]{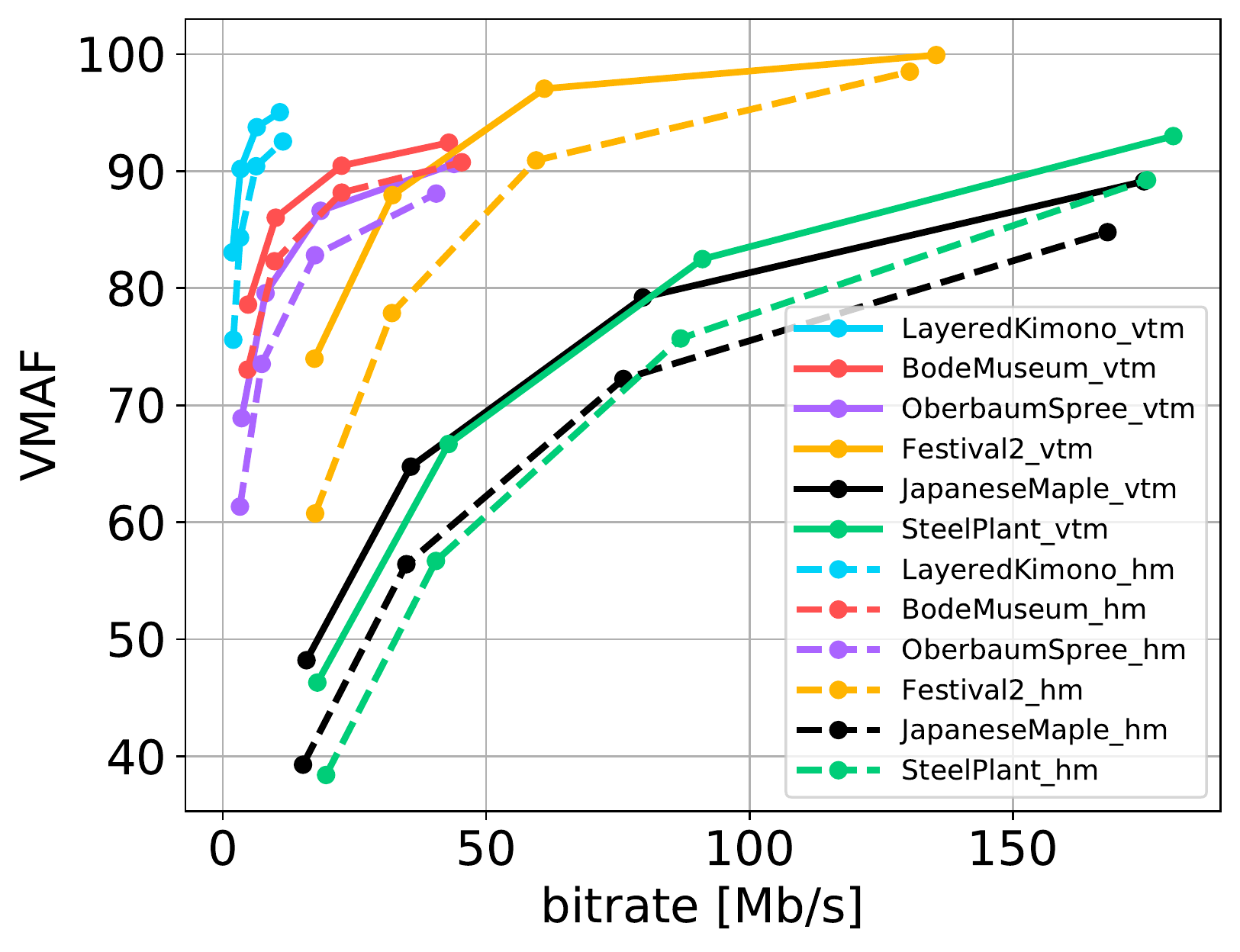}}
  \centerline{(c) VMAF}\medskip
\vspace{-0.1in}
\end{minipage}
\hfill
\caption{Objective quality comparison, using \gls{PSNR}, \acrshort{MS-SSIM}, and \gls{VMAF} quality metrics for the 8K test video sequences.}
\label{fig:obj_rd_curves}
\end{figure*}

 \begin{table*}[t]
		\caption{BD-BR scores of the \gls{VTM} codec compared to the anchor \gls{HM}. The left part of the table represents the bitrate savings (\%) for the same quality computed by objective metrics and DMOS. Negative values represent compression gain offered by \gls{VVC} over \gls{HEVC}. The right part of the table illustrates the gain in quality regarding each metric for the same bitrate. Positive values represent a gain in quality \textcolor{black}{(represented in the scale of the considered metric)} enabled by \gls{VVC} over \gls{HEVC}.}
		\label{tab:compression_perf}
		\begin{tabularx}{\linewidth}{@{}l|CCCx{4cm}|CCCx{3cm}@{}}
			\midrule[0.3mm]
			\multirow{1}{*}{Sequence}& \makecell[C]{BD-BR \\ (PSNR)} & \makecell[C]{BD-BR \\ (MS-SSIM)} & \makecell[C]{BD-BR \\ (VMAF)} & \makecell[{{x{3cm}}}]{BD-BR \\ (DMOS \textit{upper} and \textit{lower} limits)} & \makecell[C]{BD-PSNR} & \makecell[C]{BD-\\MS-SSIM} & \makecell[C]{BD-VMAF} & \makecell[{{x{3cm}}}]{BD-DMOS \\ (\textit{upper} and \textit{lower} limits)} \\
			\midrule[0.2mm]
			\multirow{1}{*}{\it LayeredKimono}& -29.77\% & -21.05\% &  -33.30\% & -44.99\% [-60.92\%, -20.04\%]& +0.61dB & +0.003 &  +4.63 & +10.76 [+19.3, +2.22]\\
			\multirow{1}{*}{\it BodeMuseum}& -32.75\% & -25.05\% & -34.70\% & -36.43\% [-74.71\%, +21.12\%] & +0.88dB & +0.002 & +3.06 & +5.79 [+15.21, -3.63]\\
			\multirow{1}{*}{\it OberbaumSpree}& -32.07\% & -27.00\% &  -33.41\% & -55.59\% [-87.15\%, +28.59\%] & +0.81dB & +0.003 &  +7.55 & +7.87 [+18.44, -3.35]\\
			\multirow{1}{*}{\it Festival2}& -36.40\% & -33.36\% &  -28.24\% & -28.89\% [-59.43\%, +37.28\%] & +1.22dB & +0.006 &  +7.37 & +5.13 [+12.98, -2.72]\\
			\multirow{1}{*}{\it JapaneseMaple}& -28.33\% & -23.37\% &  -30.86\% & -43.36\% [-64.42\%, -6.69\%] & +1.04dB & +0.009 &  +6.63 & +9.79 [+18.27, +1.31]\\
			\multirow{1}{*}{\it SteelPlant}& -28.30\% & -24.40\% & -27.57\% & -37.41\% [-67.61\%, +13.31\%] & +0.91dB & +0.007 & +7.10 & +8.83 [+20.40, -2.74]\\
			\midrule[0.2mm]
			\multirow{1}{*}{Average}&-31.27\%&-25.7\%&-35.30\%&{ -41.11\% [-69.04\%, +12.26\%]}& +0.91dB & +0.005 & +5.48 & +8.03 [+17.43, -1.49] \\
			\midrule[0.3mm]
		\end{tabularx}
\vspace{-0.1in}
\end{table*}

 In this experiment, objective quality metrics, including \gls{PSNR}, \gls{MS-SSIM}~\cite{wang2003multiscale}, and \gls{VMAF}~\cite{vmaf}, are used to measure the distortion between the 8K reconstructed signal and the source video. VMAF is an objective metric with reference, based on \gls{ML} which evaluates the quality between the source and the tested content by giving a score between 0 and 100. This metric is trained to produce a score computed from different features (motion, spatial, texture) that maximize the correlation with \gls{MOS} scores. Although \gls{VMAF} was initially optimized for visual quality estimation of 4K contents, we have integrated it into the study as it achieves a high correlation with subjective scores. In this experiment, the \gls{VMAF} scores are computed with the provided set of parameters \textit{vmaf\_v0.6.1.pkl}\footnote{\url{https://github.com/Netflix/vmaf}}. The \gls{PSNR} is assessed on the luma component only. The \gls{RD} curves are depicted in Fig. \ref{fig:obj_rd_curves}. It can be noted that the bitrates selected for transparency lead to quite different \gls{PSNR} values depending on the sequence. In contrast, for more perceptually correlated objective metrics like \gls{MS-SSIM} or \gls{VMAF}, the predicted quality converges to the maximum value for all 8K sequences. Also, those curves confirm the observation made on the scene complexity with the SI-TI graph in Fig. \ref{fig:siti}. Three categories of sequences can be distinguished by scene complexity: Group 1 includes \textit{LayeredKimono, OberbaumSpree, BodeMuseum} sequences, Group 2: \textit{Festival2}, and Group 3: \textit{JapaneseMaple, SteelPlants}. 

We use the \glsfirst{BD} computation method described in \cite{vceg_m33} to quantify the average gain in bitrate and visual quality offered by the \gls{VTM} over the \gls{HM} codec. The results are summarized in Table \ref{tab:compression_perf}. In average, the \gls{VTM} codec enables around 31\%, 26\% and 35\% of bitrate saving over the \gls{HM} codec, regarding \gls{PSNR}, \gls{MS-SSIM} and \gls{VMAF}, respectively. However, the area between the interpolated curves covered using the \gls{BD}-BR approach is limited as the selected bitrates are the same for both \gls{VVC} and \gls{HEVC}. Thus, to bring more details on the performance and consider a wider area between the curves, we compute the gain in quality of the \gls{VTM} over the \gls{HM} for the same bitrate using the \gls{BD} method. By considering this approach, 0.91dB, 0.005 and 5.48 of quality improvement is offered by the \gls{VTM} over the \gls{HM} codec for the same bitrate, regarding \gls{PSNR}, \gls{MS-SSIM} and \gls{VMAF} quality metrics, respectively.

\begin{figure*}[t]
\vspace{-0.1in}
\begin{minipage}[b]{0.33\linewidth}
  \centering
  \centerline{\includegraphics[width=1\linewidth]{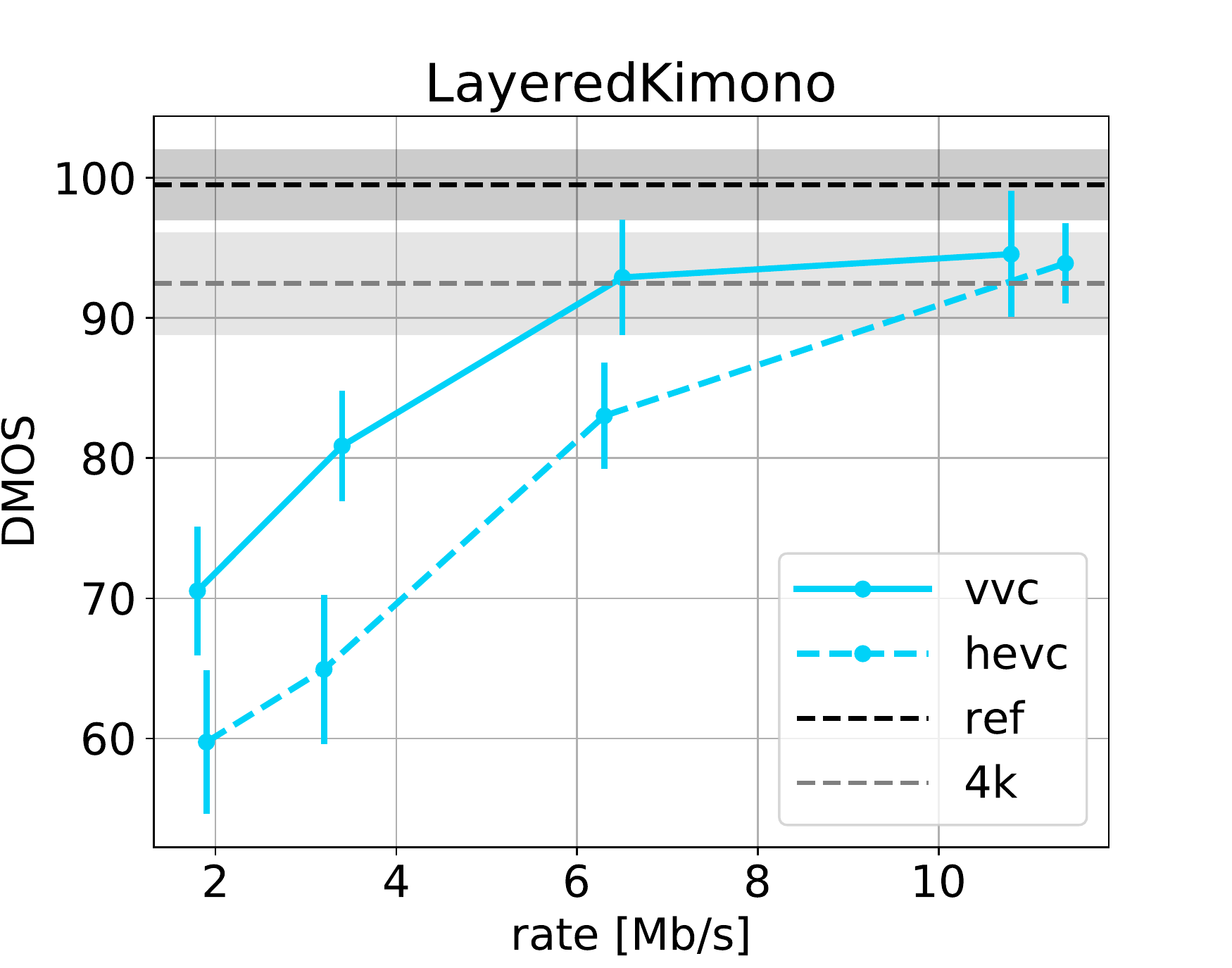}}
\end{minipage}
\begin{minipage}[b]{0.33\linewidth}
  \centering
  \centerline{\includegraphics[width=1\linewidth]{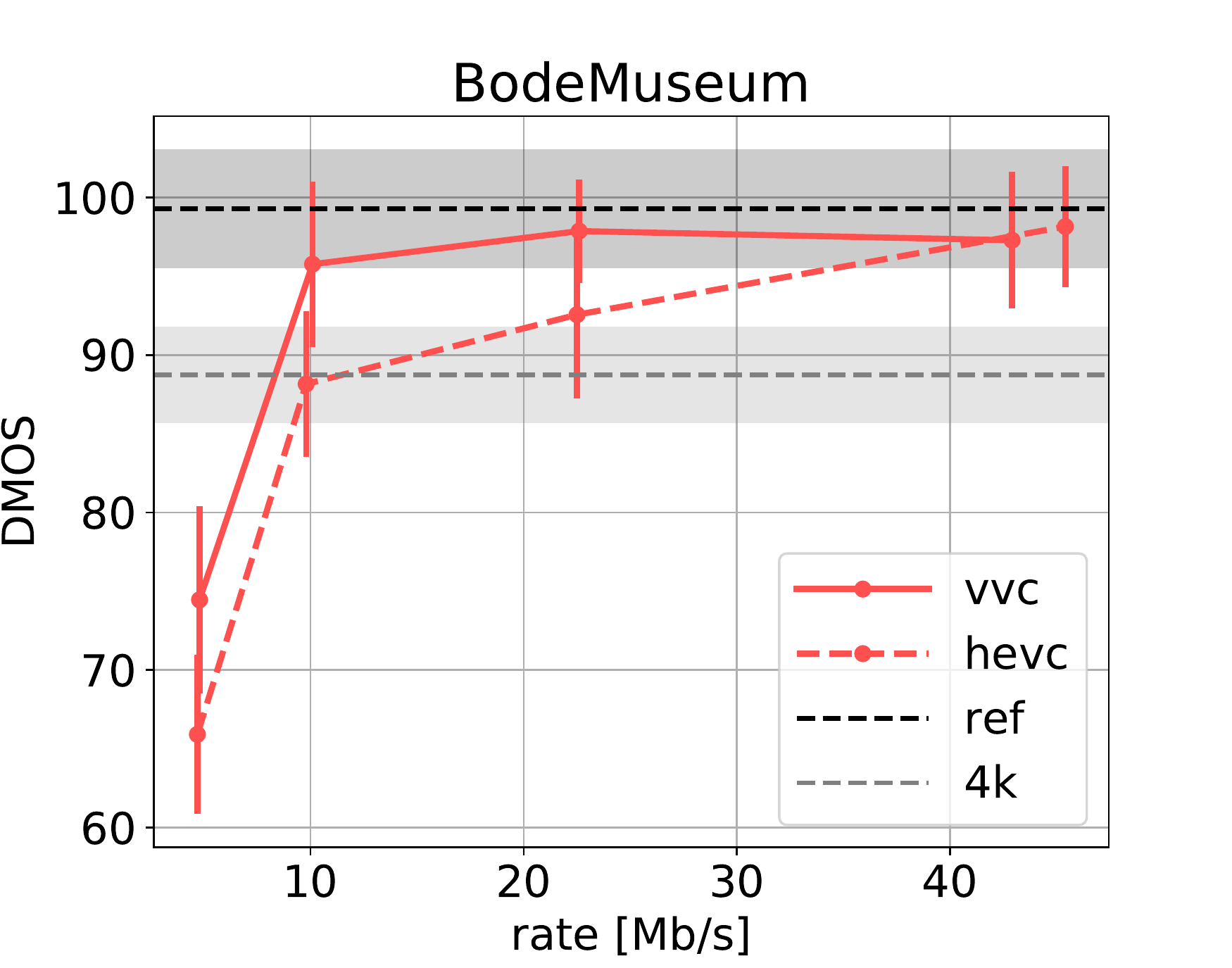}}
\end{minipage}
\hfill
\begin{minipage}[b]{0.33\linewidth}
  \centering
  \centerline{\includegraphics[width=1\linewidth]{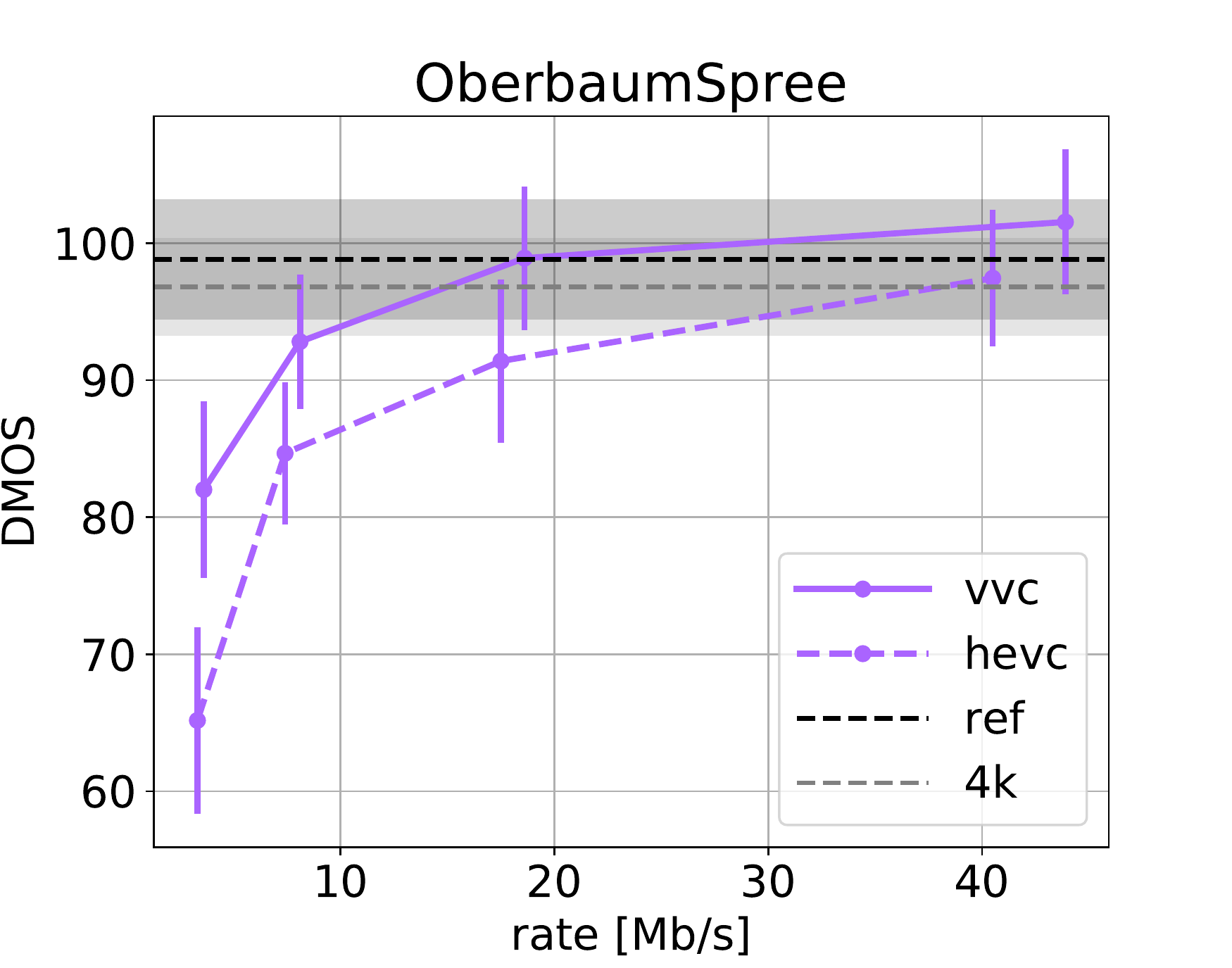}}
\end{minipage}
\hfill
\begin{minipage}[b]{0.33\linewidth}
  \centering
  \centerline{\includegraphics[width=1\linewidth]{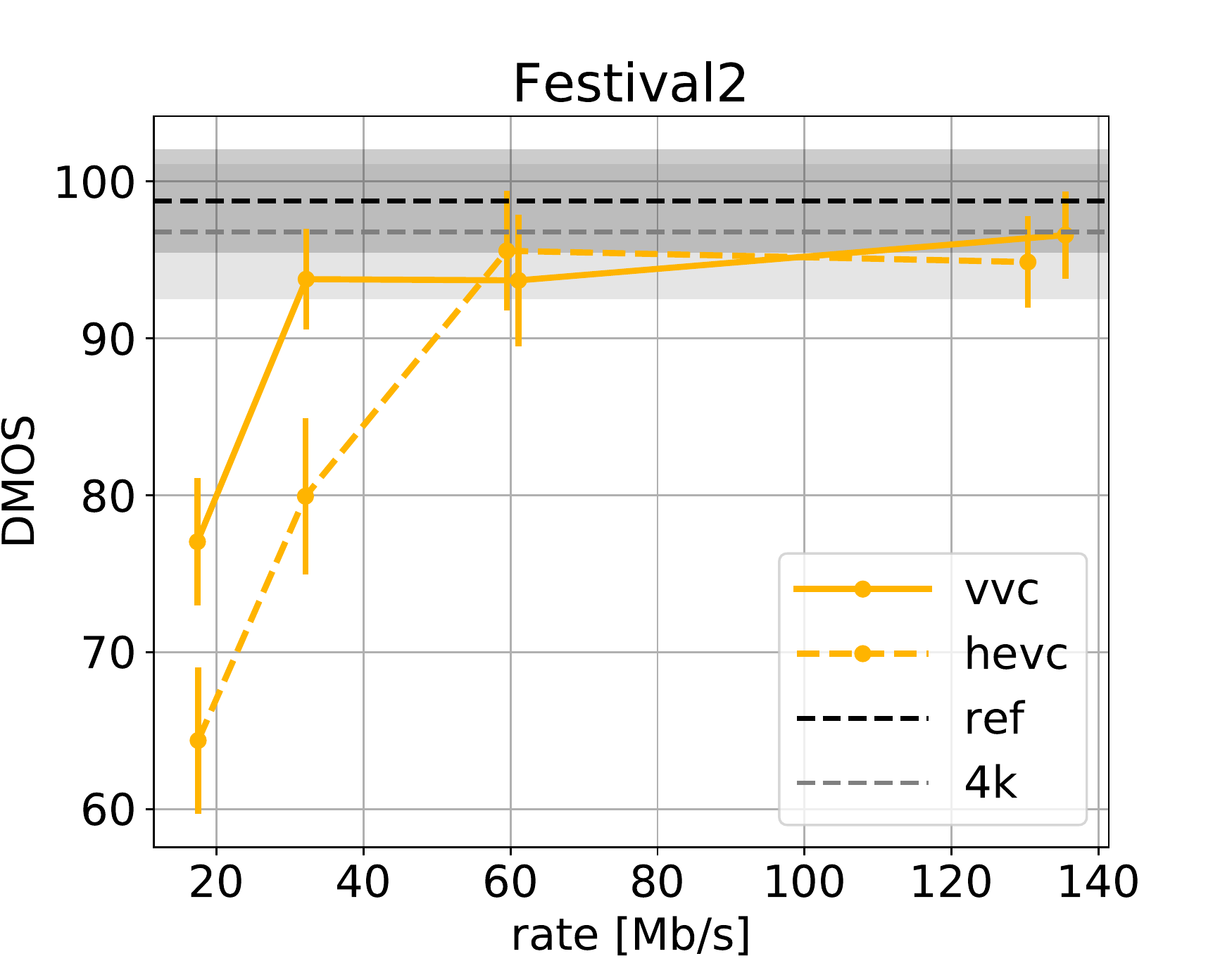}}
\vspace{-0.1in}
\end{minipage}
\begin{minipage}[b]{0.33\linewidth}
  \centering
  \centerline{\includegraphics[width=1\linewidth]{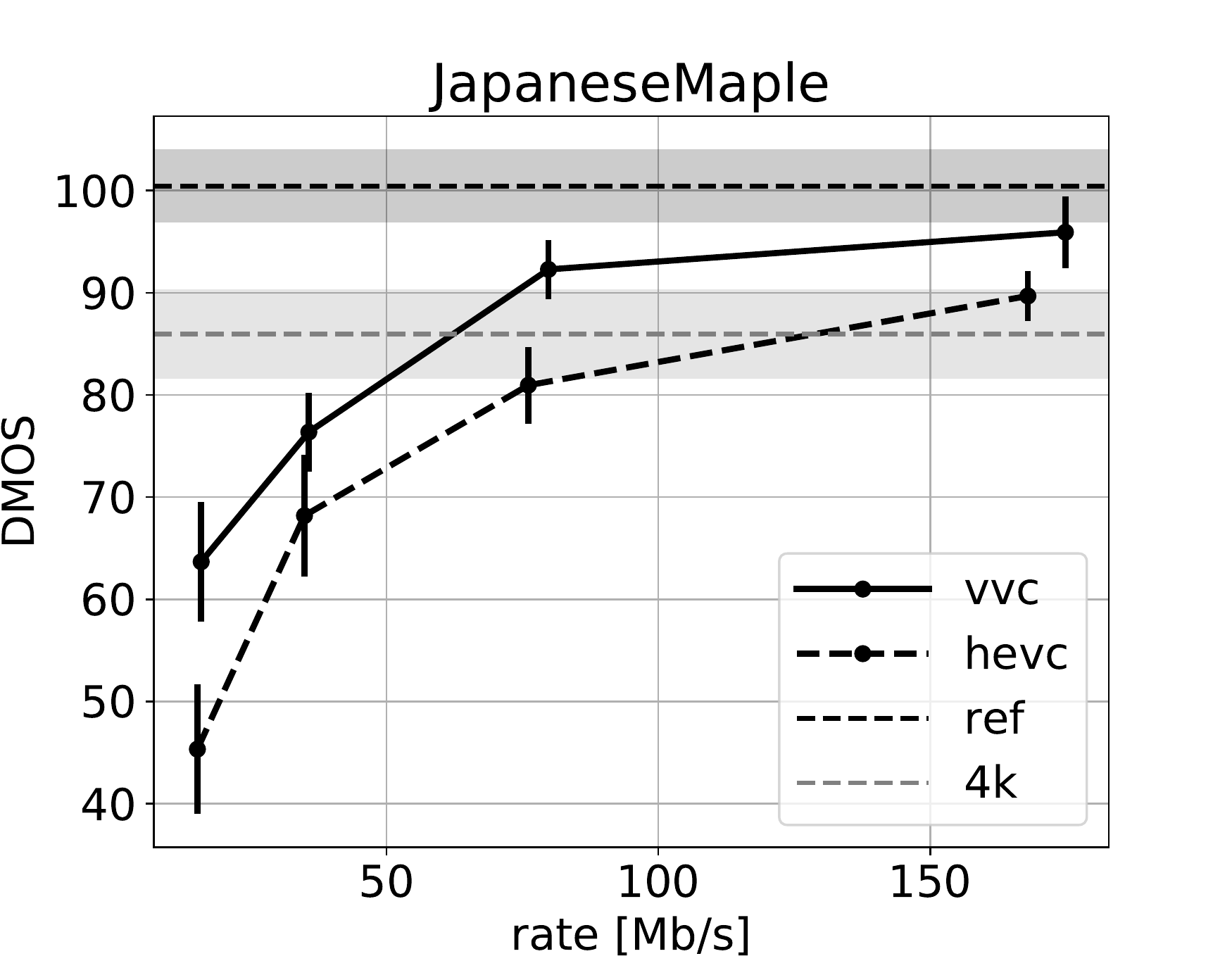}}
\vspace{-0.1in}
\end{minipage}
\hfill
\begin{minipage}[b]{0.33\linewidth}
  \centering
  \centerline{\includegraphics[width=1\linewidth]{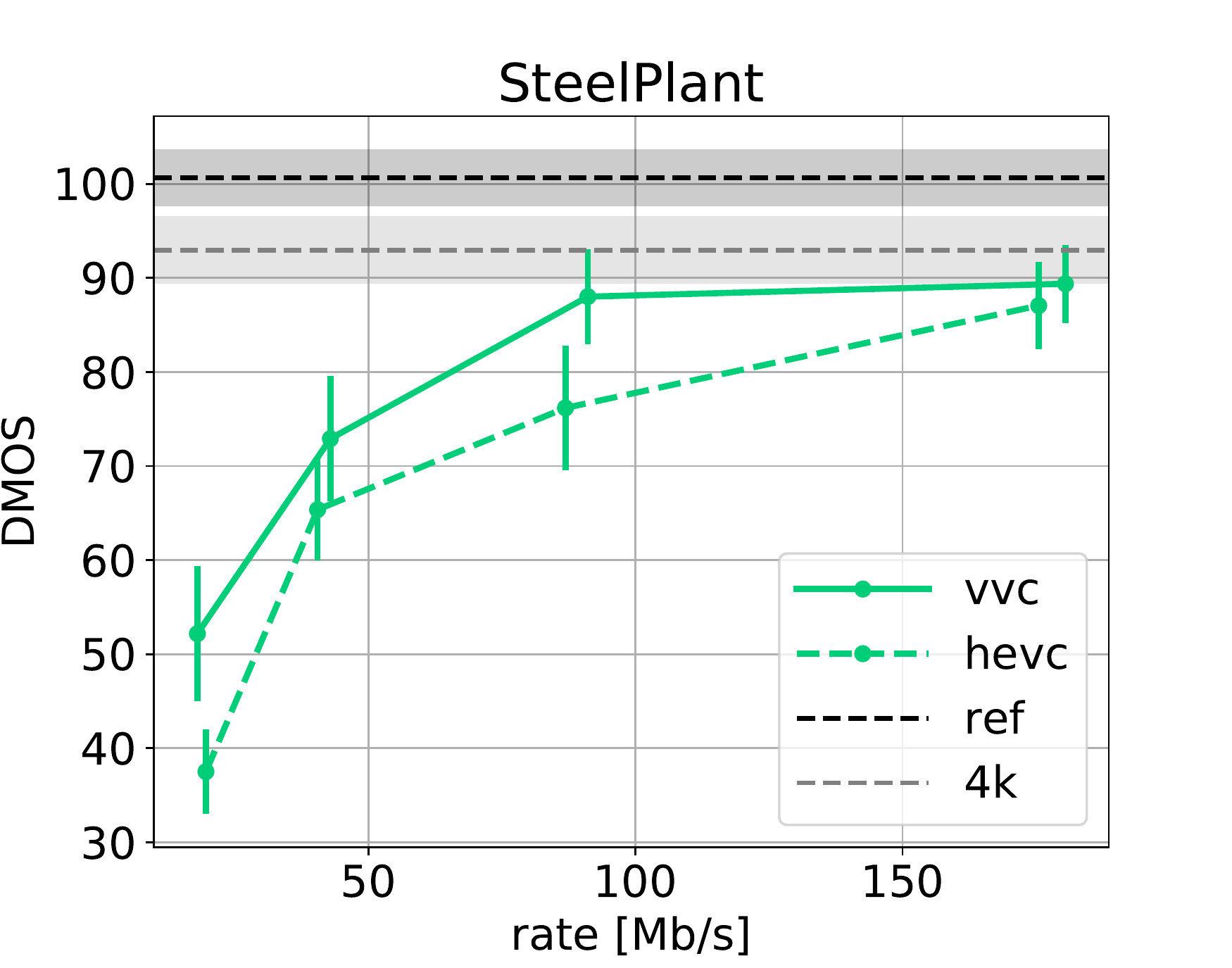}}
\vspace{-0.1in}
\end{minipage}
\hfill
\caption{DMOS-based comparison, with associated 95\% confidence interval, for the six selected 8K video sequences.}
\label{fig:rd_dmos}
\vspace{-0.1in}
\end{figure*}

\subsection{Subjective results}
\label{sec:suj_quality}

\addtolength{\tabcolsep}{-1.5pt}   
\begin{table*}[t]
  \centering
  \notsotiny
  \caption{$p$-value probabilities resulting from two-sample unequal variance bilateral Student’s t-test on DMOS values for each pair of tested configurations and each test sequence. $p \geq 0.05$ (green) means there is no significant difference between the DMOS value of the row and column labels. In contrast, $p < 0.05$ (red) indicates that the DMOS value of the row label is significantly different than the column label. \textcolor{black}{The values referred in Section~\ref{sec:suj_quality} are represented in bold.}}
  \label{tab:t-test}
  \begin{subtable}[t]{.33\linewidth}
  \caption{LayeredKimono}
  \input{tables/LayeredKimono}
  \end{subtable}
\vspace{0.1in}
  \begin{subtable}[t]{.33\linewidth}
  \caption{BodeMuseum}
    \input{tables/BodeMuseum}
  \end{subtable}
\vspace{0.1in}
  \begin{subtable}[t]{.33\linewidth}
  \caption{OberbaumSpree}
    \input{tables/OberbaumSpree}
  
  \end{subtable}
\vspace{0.1in}
  \begin{subtable}[t]{.33\linewidth}
  \caption{Festival2}
    \input{tables/Festival2}
  
  \end{subtable}
  \begin{subtable}[t]{.33\linewidth}
  \caption{JapaneseMaple}
    \input{tables/JapaneseMaple}

  \end{subtable}
  \begin{subtable}[t]{.33\linewidth}
  \caption{SteelPlant}
    \input{tables/SteelPlant}
  \end{subtable}
\vspace{-0.1in}
\end{table*}

For the subjective quality evaluation, the rectified \gls{DMOS} scores and their associated 95\% confidence interval are collected following the method described in Section \ref{sec:subj_quality_assesssment}. The resulting \gls{RD} curves are depicted in Fig. \ref{fig:rd_dmos} for all 8K sequences. These curves also display the scores obtained for the 8K hidden reference videos and the 4K sequences, with their associated 95\% confidence interval represented by transparent areas. 


In order to confidently evaluate the statistical significance of the similarity (or not) between different tested sequences, we also performed a two-sample unequal variance Student's t-test with a two-tailed distribution. This study allows us to determine, for each scene, if the perceived quality between each pair of tested configurations is significantly different or not.

In this experiment, regarding two different tested configurations $a_1$ and $a_2$ for a given scene, the null hypothesis, $H_0$, corresponds to the case that $a_1$ and $a_2$ have the same perceived quality. On the contrary, the alternate hypothesis, $H_a$, would be that a difference between the tested configurations $a_1$ and $a_2$ is noted. The t-statistic can be estimated to quantify the degree of significance of the alternate hypothesis $H_a$. By considering the sample populations $x_{a_1}$ and $x_{a_2}$ from attributed scores for the tested configuration $a_1$ and $a_2$, respectively, the t-statistic can be computed as follows:

\begin{equation}
    t_{a_1, a_2} = \frac{\bar{x}_{a_1}-\bar{x}_{a_2}}{\sqrt{\frac{s^2_{a_1}}{n_{a_1}}+\frac{s^2_{a_2}}{n_{a_2}}}},
\end{equation}

with $\bar{x}_{a_j}$, $s^2_{a_j}$ and $n_{a_j}$ denoting the mean, the variance and the size of the sample population $x_{a_j}$, with $j \in \{1,2\}$. 


Then, by approximating the t-statistic with a Student’s t-distribution, a value $p$, which indicates the degree of correlation between the means of the two sample populations, can be computed from the t-statistic. The higher the $p$-value is, the more significant the similarity between the distributions of the two populations is. A $p$-value lower than 0.05 indicates that there is a statistical significance that the two sample populations $x_{a_1}$ and $x_{a_2}$ have a different perceived quality. Indeed, there is a low probability of committing a type-I error, i.e., rejecting the null hypothesis when it is true, meaning that the null hypothesis can be confidently rejected. On the contrary, if the $p$-value is greater than or equal to 0.05, the null hypothesis cannot be safely rejected and both sample populations $x_{a_1}$ and $x_{a_2}$ can be considered to have the same perceived quality. The results for all scenes are given in Table~\ref{tab:t-test}.

The results demonstrate that the perceived quality between uncompressed 8K and 4K formats depends on the scene content. For the sequences \textit{JapaneseMaple}, \textit{SteelPlant}, \textit{BodeMuseum}, and \textit{LayeredKimono}, the visual quality between both resolutions is significantly different \textcolor{black}{as the $p$-value between the configurations 4K and REF is lower than 0.05. For those sequences, the global motion in the scene is low, which facilitate the sampling of 8K details by sensors. In contrast,} for the sequences \textit{Festival2} and \textit{OberbaumSpree}, the motion in the scene can explain the 8K definition loss at 60fps. Indeed, the global motion in \textit{Festival2} video sequence prevents from perceiving the details. For the \textit{OberbaumSpree} motion blur appears on the scene due to a continuous horizontal camera traveling. It shows that higher framerates, e.g., 100/120fps, must be considered to fully benefit from the 8K resolution. 

In complement to the objective study conducted in Section \ref{sec:obj_study}, we observe that the bitrate required to obtain transparency with the uncompressed 8K videos is highly content-dependent. Using \gls{VVC}, the bitrates needed to reach the reference's quality are between 10Mbps to 180Mbps depending on the sequence. For the \textit{SteelPlant} scene, the quality degradation with the source is always perceived on the selected bitrate range. \textcolor{black}{Indeed, the $p$-values obtained between all $R_i^{VVC}$ and REF configurations are lower than 0.05 for this sequence. It can be explained by the smoke in the scene, which is hard to compress and causes blocking artifacts.} In comparison, the 8K source quality is obtained only for three scenes using \gls{HEVC}: \textit{BodeMuseum}, \textit{Festival2}, \textit{OberbaumSpree}. However, two of them are not critical (\textit{Festival2}, \textit{OberbaumSpree}), as no significant difference between REF and 4K is perceived \textcolor{black}{$(p > 0.05)$}.

In addition, we can notice that, at the same bitrate, \gls{VVC} offers perceived quality closer to the 8K reference video comparing to \gls{HEVC}. For both \textit{JapaneseMaple} and \textit{LayeredKimono} scenes, a bitrate reduction of 50\% is reached for the same level of visual quality. Indeed, we can observe in Table~\ref{tab:t-test} that, for those two scenes, each \gls{VVC} test point of bitrate $R_i^{VVC}$ is statistically similar in terms of visual quality with respect to its corresponding \gls{HEVC} test point at bitrate $R_{i+1}^{HEVC}$ and significantly better at bitrate $R_{i}^{HEVC}$. Nevertheless, the results obtained with the rest of the 8K sequences with lower spatial textures do not follow this observation.

Finally, we applied the \gls{BD}-BR method to the \gls{DMOS} scores. Inspired by \cite{tan2015video}, we also compute the \textit{upper} and \textit{lower} limits for the \gls{BD}-BR based on the confidence intervals. These scores are computed by comparing $D^{VVC}_{max}$ with $D^{HEVC}_{min}$ and $D^{VVC}_{min}$ with $D^{HEVC}_{max}$, respectively, where $[D_{min}, D_{max}]$ represents the 95\% confidence interval. All the results are reported in Table~\ref{tab:compression_perf}. These results demonstrate that \gls{VVC} offers a compression gain over \gls{HEVC} for the same perceived quality from \textcolor{black}{28.89\% to 55.59\% with an average of 41.11\%} over the whole 8K dataset. 

\subsection{Correlation consistency}

\begin{table}[t]
		\caption{\acrshort{SROCC}, \acrshort{PLCC}, \acrshort{KROCC} and \acrshort{RMSE} performance of the objective quality metrics  \gls{MS-SSIM}, \acrshort{SSIM}, \gls{VMAF} and \gls{PSNR} on the considered 8K video sequences.}
		\label{tab:metric_perf}
		\begin{tabularx}{\linewidth}{@{}lCCCCC@{}}
			\midrule[0.3mm]
			Objective metric & SROCC & PLCC & KROCC & RMSE \\
			\midrule[0.2mm]
			\it MS-SSIM& \bfseries 0.887 & 0.871 & \bfseries 0.725 & 7.409\\
			\it SSIM& 0.767 & 0.777 & 0.599 & 9.499\\
			\it VMAF& 0.806 & \bfseries 0.873 & 0.603 & \bfseries 7.375\\
			\it PSNR& 0.754 & 0.747 & 0.564 & 10.042\\
			\midrule[0.3mm]
		\end{tabularx}
\vspace{-0.1in}
\end{table}

\begin{figure*}[t]
\begin{minipage}[b]{0.32\linewidth}
  \centering
  \centerline{\includegraphics[width=1\linewidth]{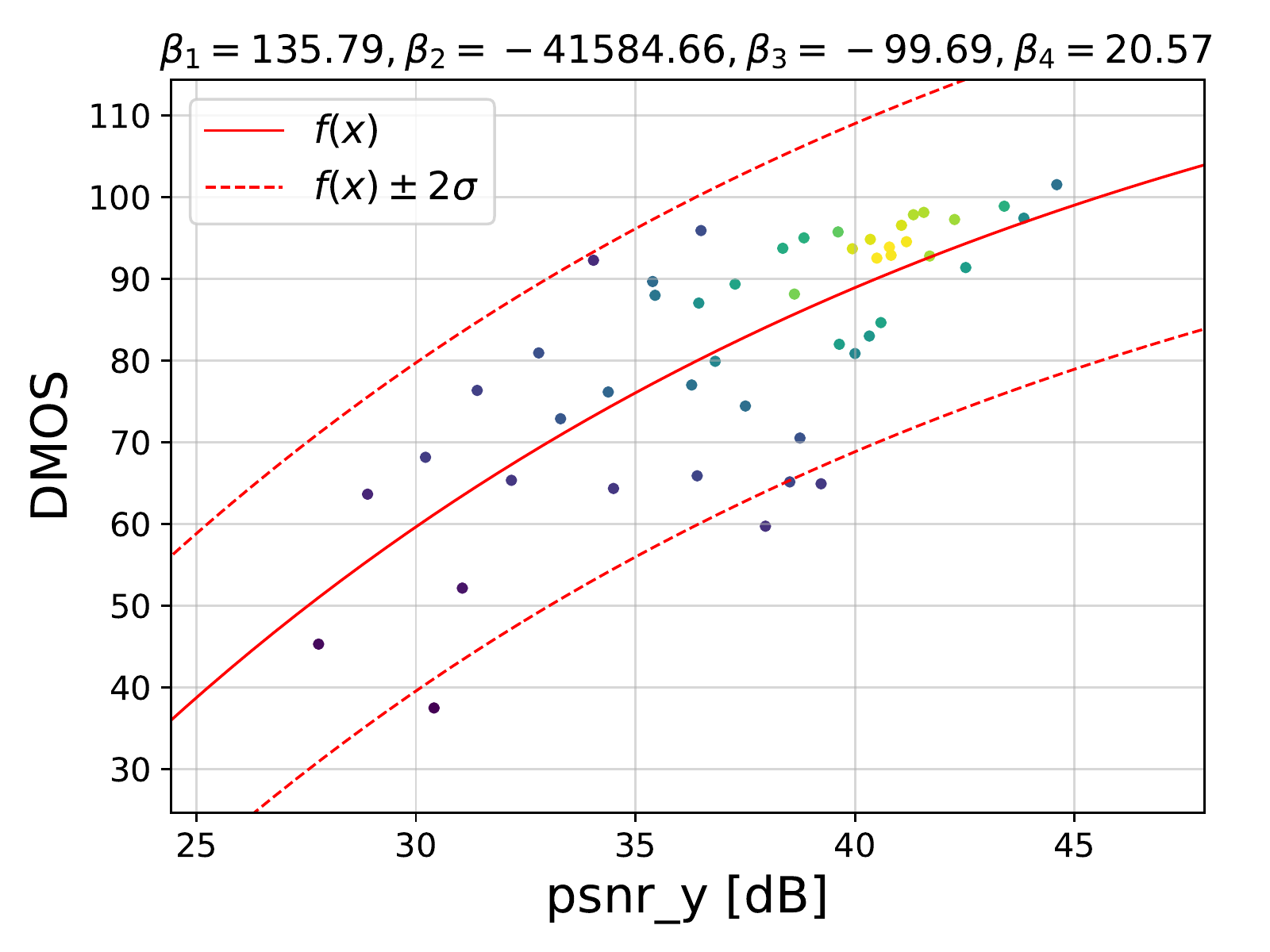}}
  \centerline{(a) PSNR}\medskip
\vspace{-0.1in}
\end{minipage}
\begin{minipage}[b]{0.32\linewidth}
  \centering
  \centerline{\includegraphics[width=1\linewidth]{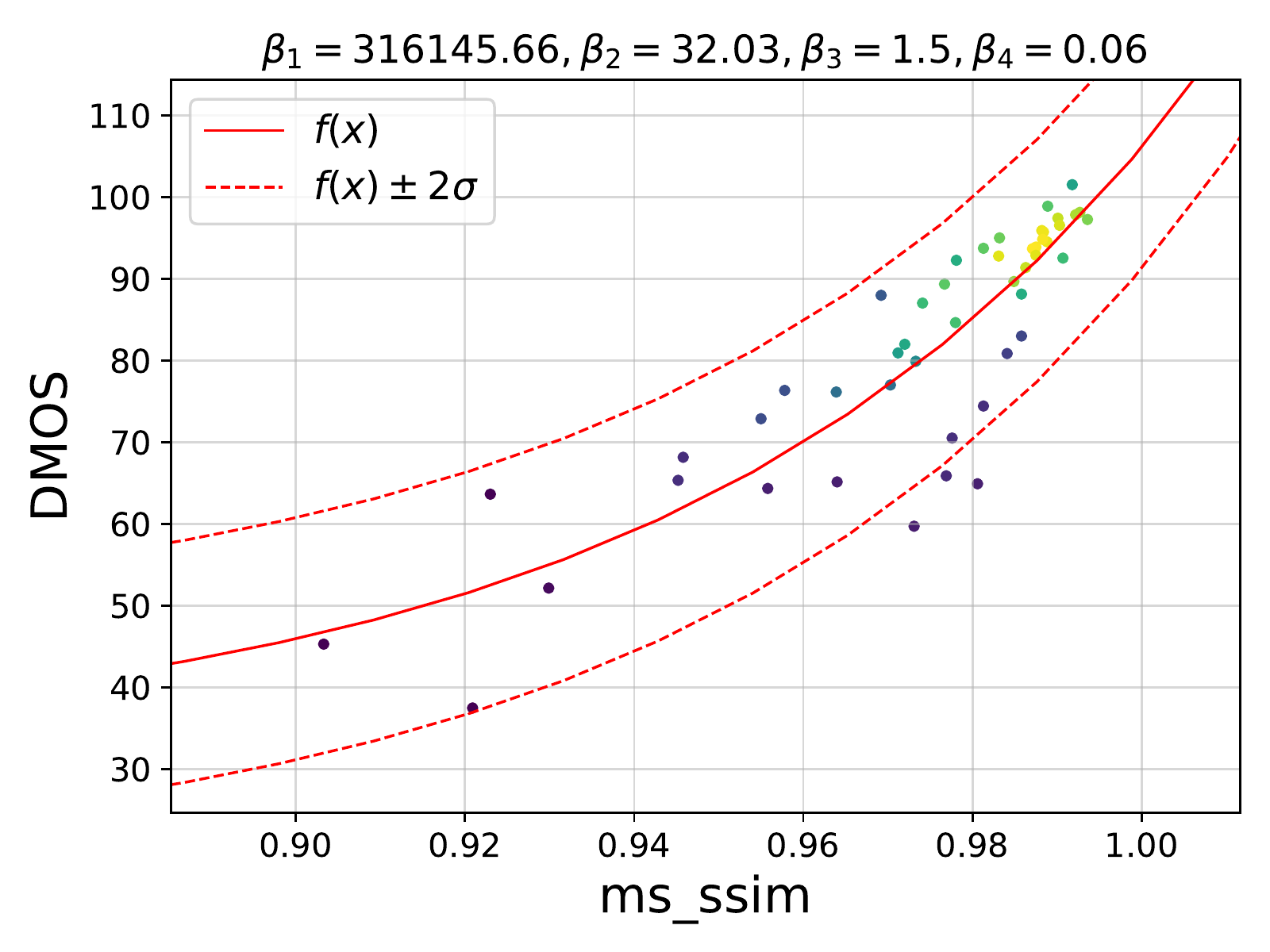}}
  \centerline{(b) MS-SSIM}\medskip
\vspace{-0.1in}
\end{minipage}
\hfill
\begin{minipage}[b]{0.32\linewidth}
  \centering
  \centerline{\includegraphics[width=1\linewidth]{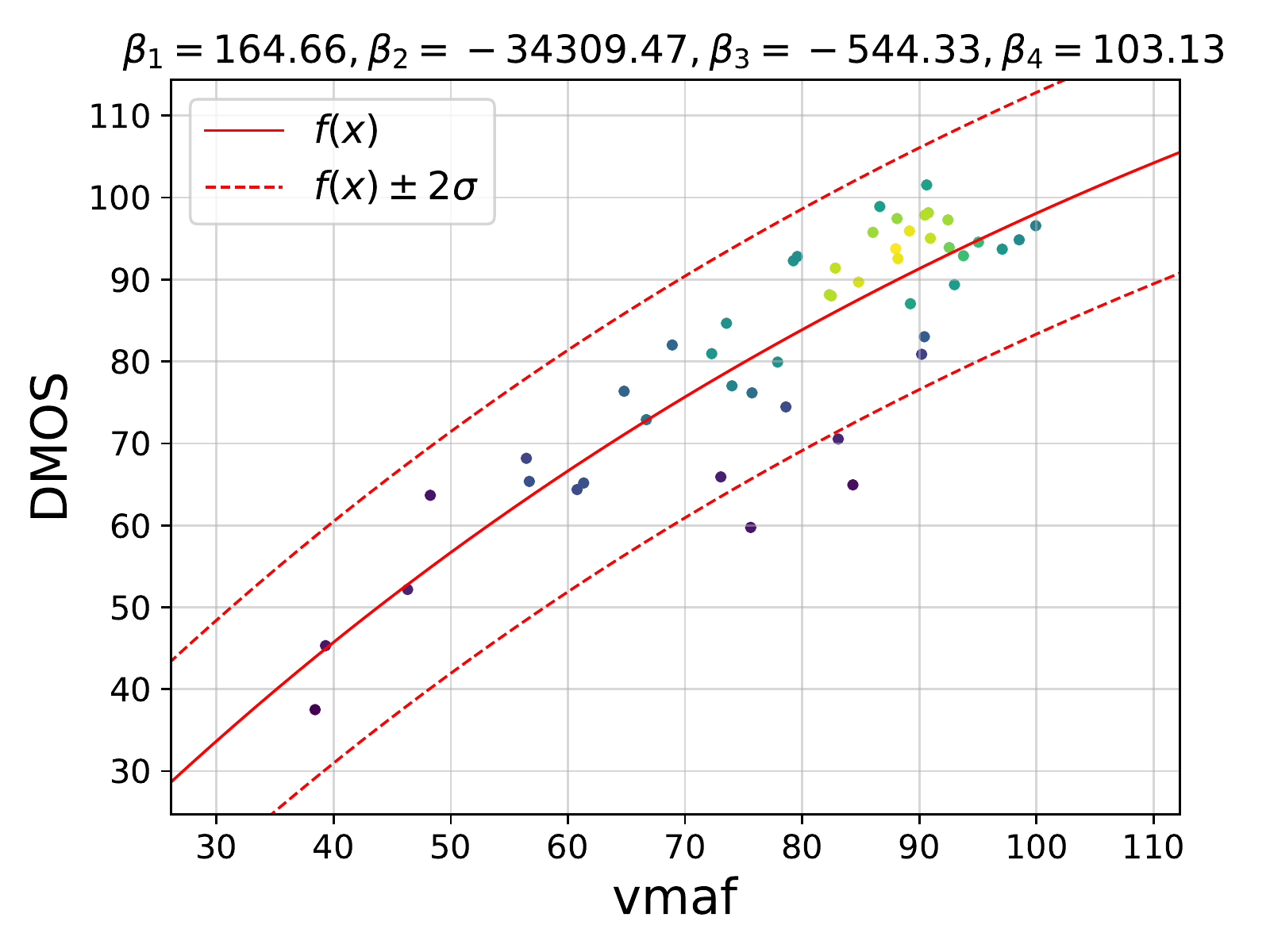}}
  \centerline{(c) VMAF}\medskip
\vspace{-0.1in}
\end{minipage}
\hfill
\caption{Scatter plots and nonlinear logistic fitted curves of PSNR, MS-SSIM and VMAF quality metrics versus \gls{DMOS} scores of the considered 8K video sequences. \textcolor{black}{The logistic model coefficients are given for each tested objective metric.}}
\label{fig:metric_perf}
\vspace{-0.2in}
\end{figure*}

In this section, the consistency of objective quality metrics with subjective scores is evaluated. Fig.~\ref{fig:metric_perf} illustrates scatter plots with nonlinear logistic fitted curves $f(x)$ and corresponding standard deviations intervals $f(x)\pm2\sigma$ for PSNR, MS-SSIM, and VMAF quality metrics versus DMOS scores. \textcolor{black}{The interpolated curves $f(x)$ are computed using the following logistic model:} 


\begin{equation}
    f(x) = \beta_2 + \frac{\beta_1-\beta_2}{1+e^{-\frac{x-\beta_3}{\abs{\beta_4}}}}.
\end{equation}

The more the standard deviation intervals are close to the logistic fitted curve, the more the metric is correlated to the \gls{DMOS} score. In order to quantify the correlation of the objective metrics with the subjective scores, we use the \glsfirst{SROCC},  \glsfirst{PLCC},  \glsfirst{KROCC}, and \glsfirst{RMSE}. The results are reported in Table~\ref{tab:metric_perf}. As expected, it shows that \gls{MS-SSIM} and \gls{VMAF} are more correlated to subjective test ratings than \gls{PSNR}, which gets the lowest performance regarding all indicators. \textcolor{black}{In addition to the three considered objective quality metrics, we provide correlation scores with the SSIM metric. This latter shows slightly higher correlation with \gls{DMOS} compared to \gls{PSNR}, while it is outperformed by both  \gls{MS-SSIM} and \gls{VMAF}.} Finally, we can notice that \gls{VMAF} is a relevant quality metric for 8K resolution evaluation although being optimized for 4K resolution.

\section{Conclusion}
\label{sec:conclusion}

In this paper, we evaluated the \gls{VVC} compression performance over its predecessor \gls{HEVC} for 8K video resolution. The subjective and objective quality assessments have been conducted on a selection of 8K video sequences in \gls{RA} configuration. Objective results have demonstrated that the \gls{VTM} codec enables 31\%, 26\% and 35\% of bitrate saving over the \gls{HM} codec, for \gls{PSNR}, \gls{MS-SSIM} and \gls{VMAF} quality metrics, respectively. On the subjective side, \gls{VVC} offers \textcolor{black}{41.11\%} of bitrate reduction over \gls{HEVC} for the same visual quality, regarding the \gls{BD}-BR method. Regarding the Student's t-test results, a bitrate reduction of about 50\% is reached for two of the overall tested scenes. We have also demonstrated that the bitrate required to obtain transparency with the 8K source is highly content-dependent. Indeed, for \gls{VVC}, a bitrate from 11Mbps to 180Mbps is needed, depending on the complexity of the scene. In addition, we demonstrated that the participants had noted a difference between uncompressed 4K and 8K for most of the tested sequences. However, sequences with high motion do not benefit from the 8K definition at 60fps. Finally, a higher correlation consistency between subjective and objective results can be noticed, particularly for the \gls{VMAF} and \gls{MS-SSIM} quality metrics. 

Future works will focus on evaluating the subjective quality offered by recent deep-learning-based tools for 8K video compression, such as super-resolution, quality enhancement, and learning-based compression methods.

\bibliographystyle{IEEEtran}
\bibliography{IEEEexample}

\begin{IEEEbiography}[{\includegraphics[width=1in,height=1.25in,clip,keepaspectratio]{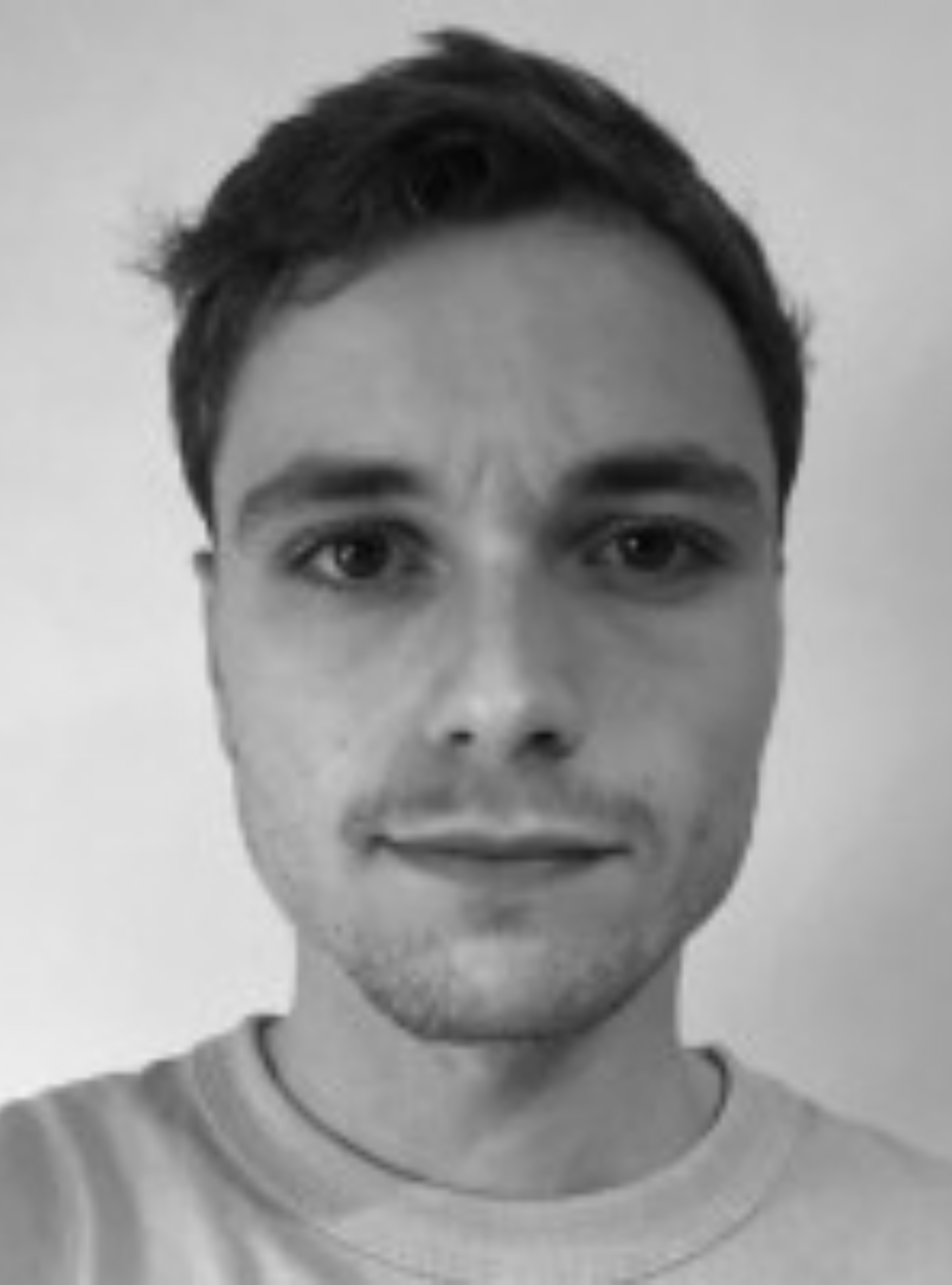}}]{Charles Bonnineau}
received the M.Sc. degree in Computer Science at the Ecole Supérieure D’Ingénieurs de Rennes (ESIR) from the University of Rennes 1, France, in 2018. He is currently a PhD Student in Signal and Image Processing jointly with the National Institute of Applied Sciences of Rennes (INSA), the Intitute of Research and Technology b$<>$com, and TDF. His principal research interests include image and video processing, learning-based post-processing algorithms for video compression, learning based image and video coding and video quality assessment.
\end{IEEEbiography}

\begin{IEEEbiography}[{\includegraphics[width=1in,height=1.25in,clip,keepaspectratio]{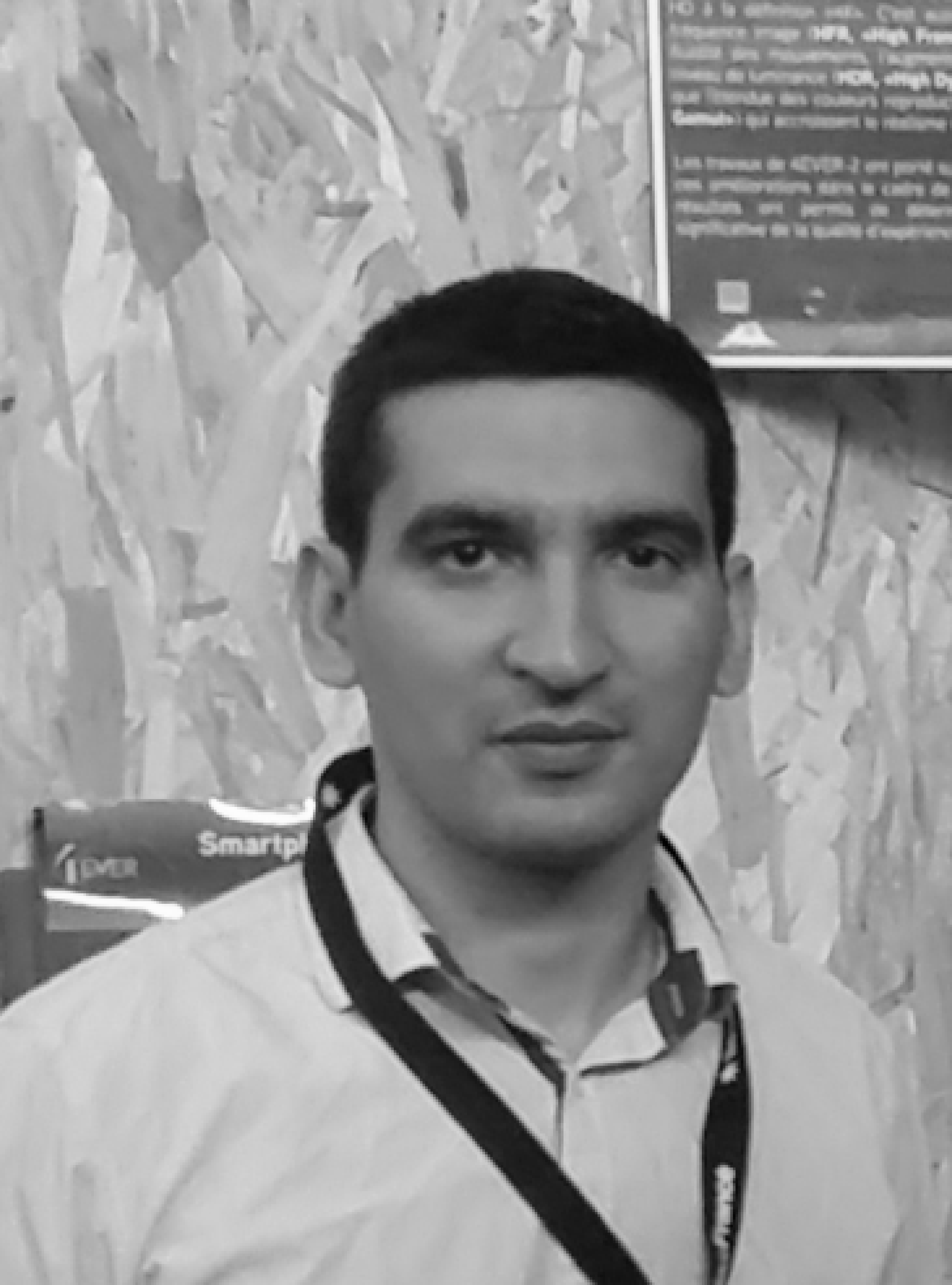}}]{Wassim Hamidouche}
received Master’s and Ph.D. degrees both in Image Processing from the University of Poitiers (France) in 2007 and 2010, respectively. From 2011 to 2013, he was a junior scientist in the video coding team of Canon Research Center in Rennes (France). He was a post-doctoral researcher from Apr. 2013 to Aug. 2015 with VAADER team of IETR where he worked under collaborative project on HEVC video standardisation. Since Sept. 2015 he is  an Associate Professor at INSA Rennes and a member of the VAADER team of IETR Lab. He has joined the Advanced Media Content Lab of b$<>$com IRT Research Institute as an academic member in Sept. 2017. His research interests focus on video coding and multimedia security. He is the author/coauthor of more than one hundred and forty papers at journals and conferences in image processing, two MPEG standards, three patents, several MPEG contributions, public datasets and open source software projects.
\end{IEEEbiography}

\begin{IEEEbiography}[{\includegraphics[width=1in,height=1.25in,clip,keepaspectratio]{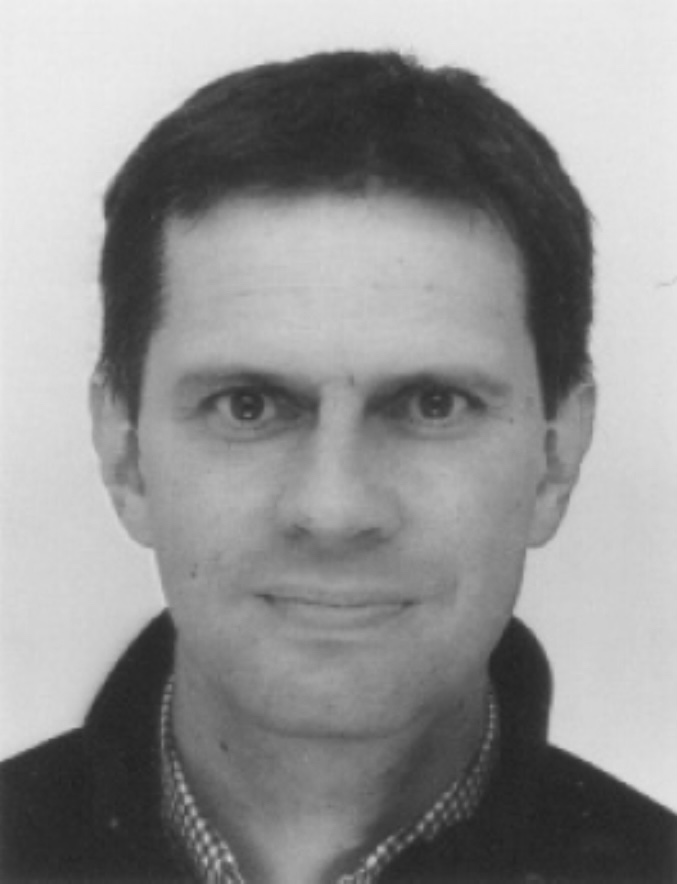}}]{Jérôme Fournier}
received the Ph.D. in signal and image processing from the University of Rennes, France, in 1995. He started his career at Philips in the field of video communications. In 1997, he joined Orange Labs (formerly France Telecom) and worked on video codecs like MPEG-4 Part 2 and H.264. From 2004 to 2012, Jérôme focused on the deployment of the Orange TV services, HDTV and stereoscopic 3DTV, as well as on innovative 3DTV depth-based video formats. From 2012 to 2018, he was mainly involved in the subjective evaluation and the ITU-R standardization of Ultra HD video formats including HDR and HFR features. Now, he is contributing to b$<>$com studies on topics like VFR, view synthesis and 8K.
\end{IEEEbiography}

\begin{IEEEbiography}[{\includegraphics[width=1in,height=1.25in,clip,keepaspectratio]{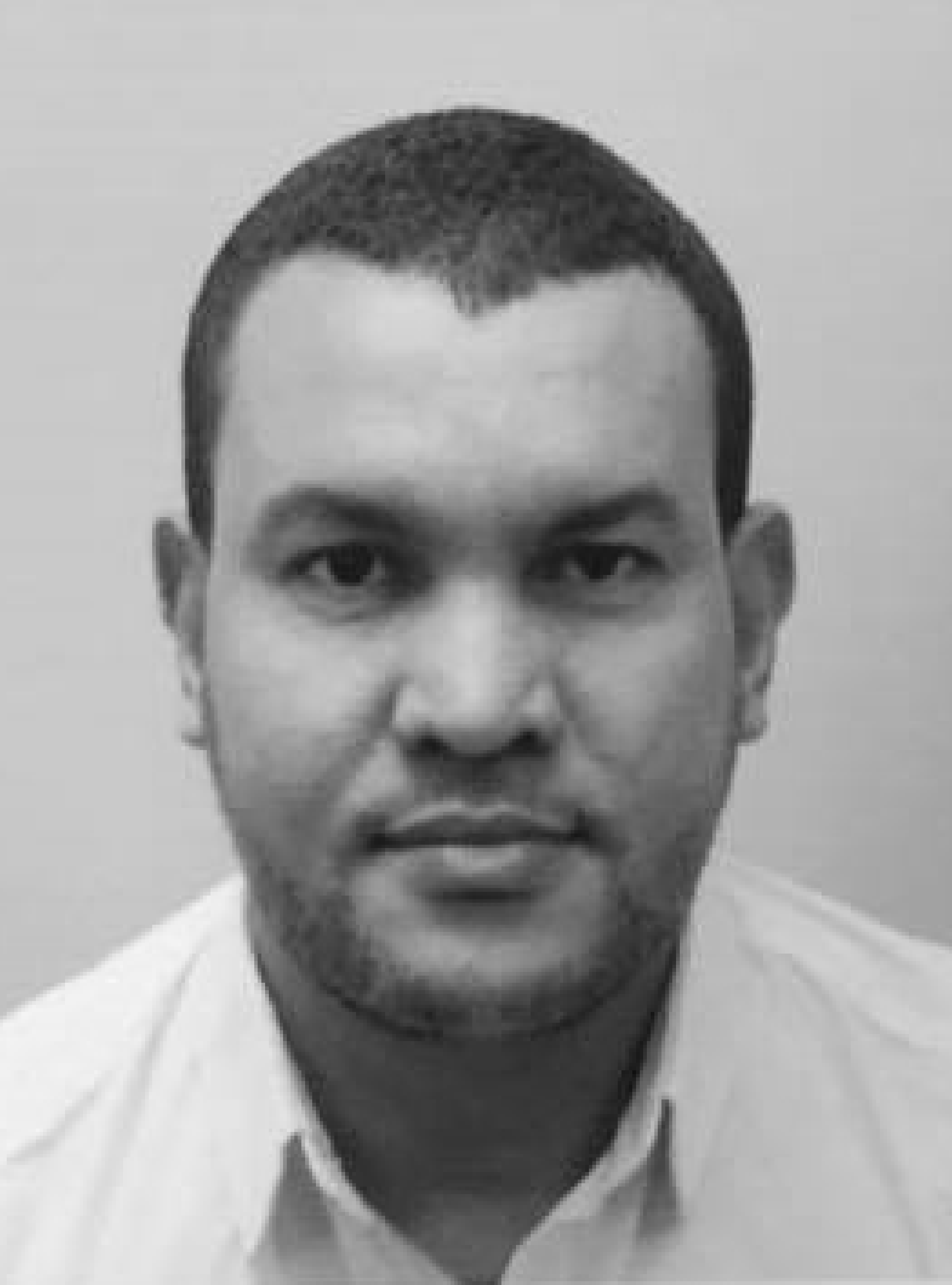}}]{Naty Sidaty}
received the Engineer and Master degrees in Telecommunications and Electronics from the National Engineering School of Tunis, Tunisia 2010, and Limoges University, France 2011, respectively. He received the Ph.D. degree in Signal and Image Processing from the University of Poitiers in 2015. From 2016 to 2019 he has been a Research Engineer with IETR Lab/INSA Rennes, Rennes, France, where he worked on the the evaluation and quality assessment of the emerging video coding standards (JEM, VVC). He is currently a Video Coding Research \& Innovation Expert at TDF group, France. He is actively involved in several standardization groups (DVB, MPEG, IUT). His research interests include Visual Attention Modeling, Video Quality Assessment, Cloud Computing, Audiovisual Services Innovation and New Formats \& Coding Tools.
\end{IEEEbiography}

\begin{IEEEbiography}[{\includegraphics[width=1in,height=1.25in,clip,keepaspectratio]{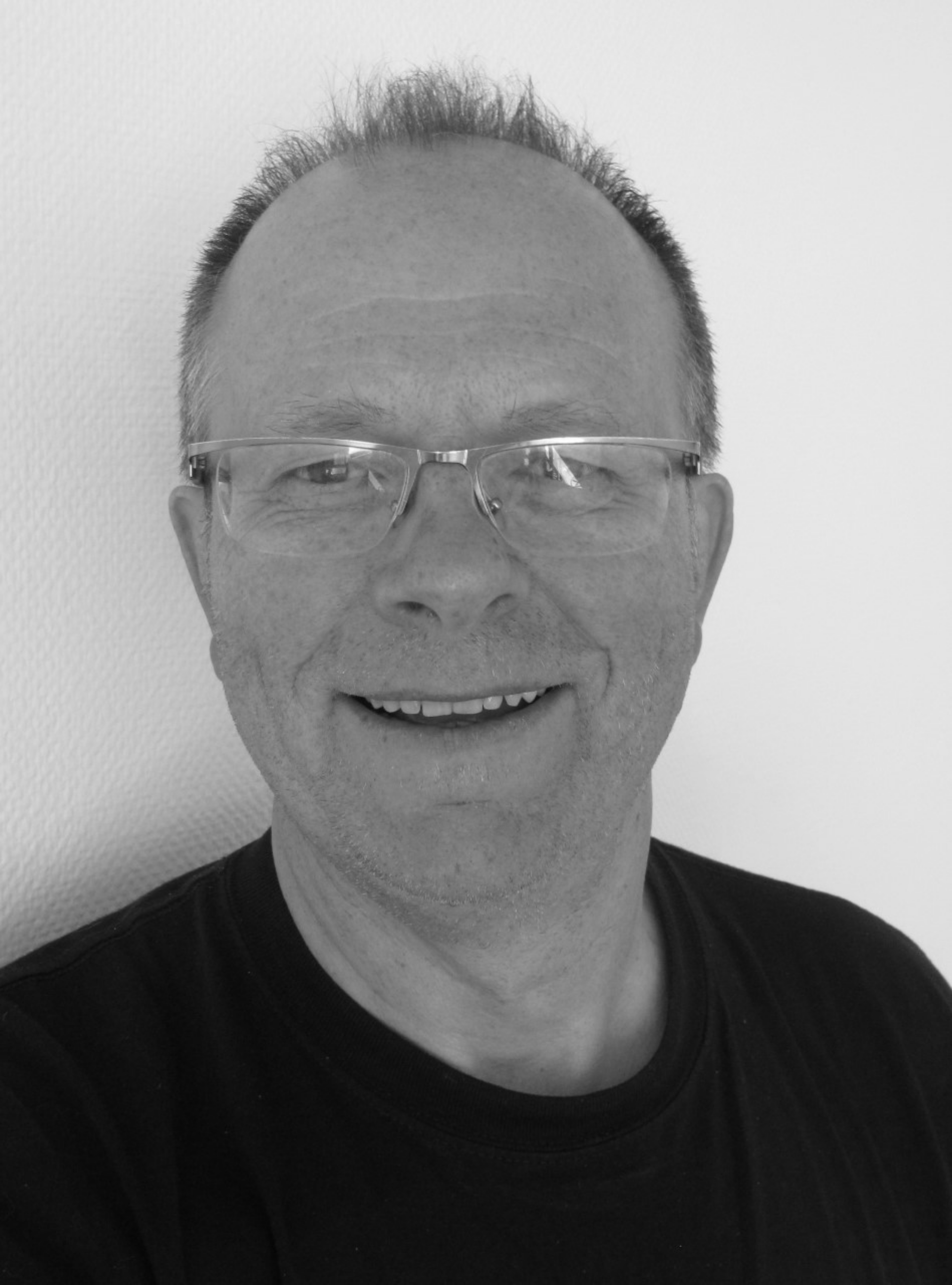}}]{Jean-François Travers}
received the M.Sc. degree in signal and image processing from the University of Rennes I, France, in 1986, and the Engineering degree in information technologies and telecommunications from ENST Bretagne in 1988. From 1989 to 1998, he was a Research Engineer with the CCETT, DAB Technologies. Since 1999, he has been an Expert in Audiovisual and System Architecture with TDF on several projects like: DTT launch and deployment, MPEG4 migration, HD and DAB+ head-ends, national and international innovative collaboration projects, and DTT UHD platform. Since 1996, he has been involved in standardization at the ETSI and DVB technical modules groups. Since 2014, project manager for TDF on French experimental DTT UHD platform and UHD demos and new services, and main contributor on new architecture of DAB+ head-end and service deployment.
\end{IEEEbiography}

\begin{IEEEbiography}[{\includegraphics[width=1in,height=1.25in,clip,keepaspectratio]{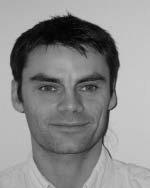}}]{Olivier Déforges}
received the Ph.D. degree in image processing, in 1995. In 1996, he joined the Department of Electronic Engineering, National Institute of Applied Sciences of Rennes (INSA), Scientic and Technical University. He is currently a Profes-
sor with INSA. He is a member of the Institute of Electronics and Telecommunications of Rennes (IETR), UMR CNRS 6164. He has authored more than 200 technical papers. His principal research interests include image and video lossy and lossless compression, image understanding, fast prototyping, and parallel architectures.
\end{IEEEbiography}

\end{document}

%% file: glossary.tex
\usepackage{hyperref}
\makeglossaries

\newacronym{VVC}{VVC}{versatile video coding}
\newacronym{HEVC}{HEVC}{high efficiency video coding}
\newacronym{QoE}{QoE}{quality of experience}
\newacronym{HDR}{HDR}{high dynamic range}
\newacronym{HFR}{HFR}{high frame-rate}
\newacronym{UHD}{UHD}{ultra high definition}
\newacronym{JVET}{JVET}{joint video exploration team}
\newacronym{ITU-T}{ITU-T}{international telecommunication union}
\newacronym{MPEG}{MPEG}{moving picture expert group}
\newacronym{AI}{AI}{artificial intelligence}
\newacronym{ANN}{ANN}{artificial neural network}
\newacronym{LR}{LR}{low-resolution}
\newacronym{HR}{HR}{high-resolution}
\newacronym{MTL}{MTL}{multitask learning}
\newacronym{QE}{QE}{quality enhancement}
\newacronym{SR}{SR}{super-resolution}
\newacronym{CR}{CR}{compact image-resolution}
\newacronym{QP}{QP}{quantization parameter}
\newacronym{VTM}{VTM-11}{VVC test model}
\newacronym{HM}{HM-16.20}{HEVC test model}
\newacronym{DSCQS}{DSCQS}{double stimulus continuous quality scale}
\newacronym{CTC}{CTC}{common test conditions}
\newacronym{BD}{BD}{Bjontegaard-Delta}
\newacronym{HD}{HD}{high-definition}
\newacronym{ML}{ML}{machine learning}
\newacronym{RB}{RB}{residual block}
\newacronym{EDSR}{EDSR}{enhanced deep super-resolution}
\newacronym{PSNR}{PSNR}{peak signal to noise ratio}
\newacronym{SSIM}{SSIM}{structural similarity}
\newacronym{BL}{BL}{base-layer}
\newacronym{EL}{EL}{enhancement-layer}
\newacronym{AE}{AE}{autoencoder}
\newacronym{HLS}{HLS}{high level syntax}
\newacronym{MSE}{MSE}{mean squared error}
\newacronym{GAN}{GAN}{generative adversarial networks}
\newacronym{SRA}{SRA}{spatial resolution adaptation}
\newacronym{CNN}{CNN}{convolutional neural network}
\newacronym{LCEVC}{LCEVC}{low complexity enhancement video coding}
\newacronym{GMM}{GMM}{gaussian mixture model}
\newacronym{GDN}{GDN}{generalized divisive normalization}
\newacronym{AE-HP}{AE-HP}{autoencoder with hyperprior}
\newacronym{CLIC21}{CLIC21}{challenge on learned image compression}
\newacronym{SHVC}{SHVC}{scalable high efficiency video coding}
\newacronym{RD}{RD}{rate-distortion}
\newacronym{bpp}{bpp}{bit per pixel}
\newacronym{ITE}{ITE}{Institute of Image Information and Television Engineers}
\newacronym{HHI}{HHI}{Fraunhofer Heinrich-Hertz-Institut}
\newacronym{RA}{RA}{random access}
\newacronym{BTC}{BTC}{basic test cell}
\newacronym{DMOS}{DMOS}{differential mean opinion score}
\newacronym{MOS}{MOS}{mean opinion score}
\newacronym{MS-SSIM}{MS-SSIM}{multi-scale structural similarity}
\newacronym{VMAF}{VMAF}{video multimethod assessment fusion}
\newacronym{PVS}{PVS}{processed video sequence}
\newacronym{SROCC}{SROCC}{Spearman’s rank ordered correlation}
\newacronym{PLCC}{PLCC}{Pearson’s linear correlation coefficient}
\newacronym{KROCC}{KROCC}{Kendall’s rank-order correlation coefficient}
\newacronym{RMSE}{RMSE}{root mean-squared error}
\newacronym{AVC}{AVC}{advanced video coding}
\newacronym{UHDTV}{UHDTV}{ultra-high definition television}
\newacronym{HDTV}{HDTV}{high definition television}
\newacronym{UHF}{UHF}{ultra high frequency}
\newacronym{AV1}{AV1}{AOMedia Video 1}

%% file: tables/LayeredKimono.tex
\begin{tabular}[t]{l|rrrr|rr}
\diagbox[width=\dimexpr \textwidth/6+2\tabcolsep\relax, height=0.8cm]{HEVC}{VVC} &  R1 &  R2 &  R3 &  R4 &    4K &   REF \\
\hline
R1 &    \bfseries\cellcolor{redtable}{0.01} &    \cellcolor{redtable}{0.00} &    \cellcolor{redtable}{0.00} &    \cellcolor{redtable}{0.00} &   \cellcolor{redtable}{0.00} &   \bfseries\cellcolor{redtable}{0.00} \\
R2 &    \bfseries\cellcolor{greentable}{0.15} &    \bfseries\cellcolor{redtable}{0.00} &    \cellcolor{redtable}{0.00} &    \cellcolor{redtable}{0.00} &   \cellcolor{redtable}{0.00} &   \bfseries\cellcolor{redtable}{0.00} \\
R3 &    \cellcolor{redtable}{0.00} &    \bfseries\cellcolor{greentable}{0.44} &    \bfseries\cellcolor{redtable}{0.00} &    \cellcolor{redtable}{0.00} &   \cellcolor{redtable}{0.00} &   \bfseries\cellcolor{redtable}{0.00} \\
R4 &    \cellcolor{redtable}{0.00} &    \cellcolor{redtable}{0.00} &    \bfseries\cellcolor{greentable}{0.79} &    \bfseries\cellcolor{greentable}{0.70} &   \cellcolor{greentable}{0.65} &   \bfseries\cellcolor{redtable}{0.01} \\
\hline
4K     &    \cellcolor{redtable}{0.00} &    \cellcolor{redtable}{0.00} &    \cellcolor{greentable}{0.88} &    \cellcolor{greentable}{0.47} &   \bfseries\cellcolor{greentable}{1.00} &   \bfseries\cellcolor{redtable}{0.01} \\
REF    &    \bfseries\cellcolor{redtable}{0.00} &    \bfseries\cellcolor{redtable}{0.00} &    \bfseries\cellcolor{redtable}{0.02} &    \bfseries\cellcolor{greentable}{0.10} &   \bfseries\cellcolor{redtable}{0.01} &   \bfseries\cellcolor{greentable}{1.00} \\
\hline
\end{tabular}

%% file: tables/BodeMuseum.tex
\begin{tabular}[t]{l|rrrr|rr}
\diagbox[width=\dimexpr \textwidth/6+2\tabcolsep\relax, height=0.8cm]{HEVC}{VVC} &  R1 &  R2 &  R3 &  R4 &    4K &   REF \\
\hline
R1 &   \bfseries\cellcolor{redtable}{0.04} &   \cellcolor{redtable}{0.00} & \cellcolor{redtable}{0.00} &   \cellcolor{redtable}{0.00} &  \cellcolor{redtable}{0.00} &  \bfseries\cellcolor{redtable}{0.00} \\
R2 &   \bfseries\cellcolor{redtable}{0.00} &   \bfseries\cellcolor{greentable}{0.06} &   \cellcolor{redtable}{0.00} &   \cellcolor{redtable}{0.01} &  \cellcolor{greentable}{0.90} &  \bfseries\cellcolor{redtable}{0.00} \\
R3 &   \cellcolor{redtable}{0.00} &   \bfseries\cellcolor{greentable}{0.44} &   \bfseries\cellcolor{greentable}{0.13} &   \cellcolor{greentable}{0.21} &  \cellcolor{greentable}{0.28} &  \bfseries\cellcolor{greentable}{0.07} \\
R4 &   \cellcolor{redtable}{0.00} &   \cellcolor{greentable}{0.56} &   \bfseries\cellcolor{greentable}{0.98} &   \bfseries\cellcolor{greentable}{0.86} &  \cellcolor{redtable}{0.00} &  \bfseries\cellcolor{greentable}{0.62} \\
\hline
4K     &   \cellcolor{redtable}{0.00} &   \cellcolor{greentable}{0.05} &   \cellcolor{redtable}{0.00} &   \cellcolor{redtable}{0.01} &  \bfseries\cellcolor{greentable}{1.00} &  \bfseries\cellcolor{redtable}{0.00} \\
REF    &   \bfseries\cellcolor{redtable}{0.00} &   \bfseries\cellcolor{greentable}{0.32} &   \bfseries\cellcolor{greentable}{0.58} &   \bfseries\cellcolor{greentable}{0.53} &  \bfseries\cellcolor{redtable}{0.00} &  \bfseries\cellcolor{greentable}{1.00} \\
\hline
\end{tabular}

%% file: tables/OberbaumSpree.tex
\begin{tabular}[t]{l|rrrr|rr}
\diagbox[width=\dimexpr \textwidth/6+2\tabcolsep\relax, height=0.8cm]{HEVC}{VVC} &  R1 &  R2 &  R3 &  R4 &    4K &   REF \\
\hline
R1 &    \bfseries\cellcolor{redtable}{0.00} &    \cellcolor{redtable}{0.00} &    \cellcolor{redtable}{0.00} &    \cellcolor{redtable}{0.00} &   \cellcolor{redtable}{0.00} &   \bfseries\cellcolor{redtable}{0.00} \\
R2 &    \bfseries\cellcolor{greentable}{0.61} &    \bfseries\cellcolor{redtable}{0.04} &    \cellcolor{redtable}{0.00} &    \cellcolor{redtable}{0.00} &   \cellcolor{redtable}{0.04} &   \bfseries\cellcolor{redtable}{0.01} \\
R3 &    \cellcolor{greentable}{0.06} &    \bfseries\cellcolor{greentable}{0.74} &    \bfseries\cellcolor{greentable}{0.09} &    \cellcolor{redtable}{0.02} &   \cellcolor{greentable}{0.16} &   \bfseries\cellcolor{greentable}{0.07} \\
R4 &    \cellcolor{redtable}{0.00} &    \cellcolor{greentable}{0.23} &    \bfseries\cellcolor{greentable}{0.71} &    \bfseries\cellcolor{greentable}{0.31} &   \cellcolor{greentable}{0.85} &   \bfseries\cellcolor{greentable}{0.71} \\
\hline
4K     &    \cellcolor{redtable}{0.00} &    \cellcolor{greentable}{0.23} &    \cellcolor{greentable}{0.55} &    \cellcolor{greentable}{0.18} &   \bfseries\cellcolor{greentable}{1.00} &   \bfseries\cellcolor{greentable}{0.52} \\
REF    &    \bfseries\cellcolor{redtable}{0.00} &    \bfseries\cellcolor{greentable}{0.10} &    \bfseries\cellcolor{greentable}{0.98} &    \bfseries\cellcolor{greentable}{0.47} &   \bfseries\cellcolor{greentable}{0.52} &   \bfseries\cellcolor{greentable}{1.00} \\
\hline
\end{tabular}

%% file: tables/Festival2.tex
\begin{tabular}[t]{l|rrrr|rr}
\diagbox[width=\dimexpr \textwidth/6+2\tabcolsep\relax, height=0.8cm]{HEVC}{VVC} &  R1 &  R2 &  R3 &  R4 &    4K &   REF \\
\hline
R1 &    \bfseries\cellcolor{redtable}{0.00} &    \cellcolor{redtable}{0.00} &    \cellcolor{redtable}{0.00} &    \cellcolor{redtable}{0.00} &   \cellcolor{redtable}{0.00} &   \bfseries\cellcolor{redtable}{0.00} \\
R2 &    \bfseries\cellcolor{greentable}{0.41} &    \bfseries\cellcolor{redtable}{0.00} &    \cellcolor{redtable}{0.00} &    \cellcolor{redtable}{0.00} &   \cellcolor{redtable}{0.00} &   \bfseries\cellcolor{redtable}{0.00} \\
R3 &    \cellcolor{redtable}{0.00} &    \bfseries\cellcolor{greentable}{0.34} &    \bfseries\cellcolor{greentable}{0.55} &    \cellcolor{greentable}{0.70} &   \cellcolor{greentable}{0.73} &   \bfseries\cellcolor{greentable}{0.26} \\
R4 &    \cellcolor{redtable}{0.00} &    \cellcolor{greentable}{0.42} &    \bfseries\cellcolor{greentable}{0.68} &    \bfseries\cellcolor{greentable}{0.44} &   \cellcolor{greentable}{0.53} &   \bfseries\cellcolor{greentable}{0.11} \\
\hline
4K     &    \cellcolor{redtable}{0.00} &    \cellcolor{greentable}{0.21} &    \cellcolor{greentable}{0.37} &    \cellcolor{greentable}{0.98} &   \bfseries\cellcolor{greentable}{1.00} &   \bfseries\cellcolor{greentable}{0.48} \\
REF    &    \bfseries\cellcolor{redtable}{0.00} &    \bfseries\cellcolor{redtable}{0.02} &    \bfseries\cellcolor{greentable}{0.09} &    \bfseries\cellcolor{greentable}{0.36} &   \bfseries\cellcolor{greentable}{0.48} &   \bfseries\cellcolor{greentable}{1.00} \\
\hline
\end{tabular}

%% file: tables/JapaneseMaple.tex
\begin{tabular}[t]{l|rrrr|rr}
\diagbox[width=\dimexpr \textwidth/6+2\tabcolsep\relax, height=0.8cm]{HEVC}{VVC} &  R1 &  R2 &  R3 &  R4 &    4K &   REF \\
\hline
R1 &    \bfseries\cellcolor{redtable}{0.00} &    \cellcolor{redtable}{0.00} &    \cellcolor{redtable}{0.00} &    \cellcolor{redtable}{0.00} &   \cellcolor{redtable}{0.00} &   \bfseries\cellcolor{redtable}{0.00} \\
R2 &    \bfseries\cellcolor{greentable}{0.33} &    \bfseries\cellcolor{redtable}{0.04} &    \cellcolor{redtable}{0.00} &    \cellcolor{redtable}{0.00} &   \cellcolor{redtable}{0.00} &   \bfseries\cellcolor{redtable}{0.00} \\
R3 &    \cellcolor{redtable}{0.00} &    \bfseries\cellcolor{greentable}{0.13} &    \bfseries\cellcolor{redtable}{0.00} &    \cellcolor{redtable}{0.00} &   \cellcolor{greentable}{0.12} &   \bfseries\cellcolor{redtable}{0.00} \\
R4 &    \cellcolor{redtable}{0.00} &    \cellcolor{redtable}{0.00} &    \bfseries\cellcolor{greentable}{0.24} &    \bfseries\cellcolor{redtable}{0.00} &   \cellcolor{greentable}{0.18} &   \bfseries\cellcolor{redtable}{0.00} \\
\hline
4K     &    \cellcolor{redtable}{0.00} &    \cellcolor{redtable}{0.00} &    \cellcolor{redtable}{0.04} &    \cellcolor{redtable}{0.00} &   \bfseries\cellcolor{greentable}{1.00} &   \bfseries\cellcolor{redtable}{0.00} \\
REF    &    \bfseries\cellcolor{redtable}{0.00} &    \bfseries\cellcolor{redtable}{0.00} &    \bfseries\cellcolor{redtable}{0.00} &    \bfseries\cellcolor{greentable}{0.14} &   \bfseries\cellcolor{redtable}{0.00} &   \bfseries\cellcolor{greentable}{1.00} \\
\hline
\end{tabular}

%% file: tables/SteelPlant.tex
\begin{tabular}[t]{l|rrrr|rr}
\diagbox[width=\dimexpr \textwidth/6+2\tabcolsep\relax, height=0.8cm]{HEVC}{VVC} &  R1 &  R2 &  R3 &  R4 &    4K &   REF \\
\hline
R1 &    \bfseries\cellcolor{redtable}{0.00} &    \cellcolor{redtable}{0.00} &    \cellcolor{redtable}{0.00} &    \cellcolor{redtable}{0.00} &   \cellcolor{redtable}{0.00} &   \bfseries\cellcolor{redtable}{0.00} \\
R2 &    \bfseries\cellcolor{redtable}{0.00} &    \bfseries\cellcolor{greentable}{0.11} &    \cellcolor{redtable}{0.00} &    \cellcolor{redtable}{0.00} &   \cellcolor{redtable}{0.00} &   \bfseries\cellcolor{redtable}{0.00} \\
R3 &    \cellcolor{redtable}{0.00} &    \bfseries\cellcolor{greentable}{0.55} &    \bfseries\cellcolor{redtable}{0.01} &    \cellcolor{redtable}{0.00} &   \cellcolor{redtable}{0.00} &   \bfseries\cellcolor{redtable}{0.0} \\
R4 &    \cellcolor{redtable}{0.00} &    \cellcolor{redtable}{0.00} &    \bfseries\cellcolor{greentable}{0.91} &    \bfseries\cellcolor{greentable}{0.50} &   \cellcolor{greentable}{0.07} &   \bfseries\cellcolor{redtable}{0.00} \\
\hline
4K     &    \cellcolor{redtable}{0.00} &    \cellcolor{redtable}{0.00} &    \cellcolor{greentable}{0.11} &    \cellcolor{greentable}{0.24} &   \bfseries\cellcolor{greentable}{1.00} &   \bfseries\cellcolor{redtable}{0.00} \\
REF    &    \bfseries\cellcolor{redtable}{0.00} &    \bfseries\cellcolor{redtable}{0.00} &    \bfseries\cellcolor{redtable}{0.00} &    \bfseries\cellcolor{redtable}{0.00} &   \bfseries\cellcolor{redtable}{0.00} &   \bfseries\cellcolor{greentable}{1.00} \\
\hline
\end{tabular}